\documentclass[aps,prc,twocolumn,superscriptaddress,floatfix]{revtex4}

\usepackage{amsmath}
\usepackage[english]{babel}
\usepackage[latin1]{inputenc}
\usepackage{amssymb}
\usepackage{graphicx}
\usepackage{natbib}
\usepackage{epsfig}
\usepackage{bm}
\usepackage{dcolumn}
\usepackage{color}

\def\openone{\leavevmode\hbox{\small1\kern-3.8pt\normalsize1}}%

\def\mf{{\mbox{\tiny\em MFA}}}

\def\bea{\begin{eqnarray}}
\def\eea{\end{eqnarray}}
\def\beq{\begin{equation}}
\def\eeq{\end{equation}}

\begin{document}

\title{Third family of compact stars within a nonlocal chiral quark model equation of state}

\author{D.~E.~Alvarez-Castillo}
\email{alvarez@theor.jinr.ru}
\affiliation{Bogoliubov  Laboratory of Theoretical Physics,
	Joint Institute for Nuclear Research,\\
	Joliot-Curie str. 6, 141980 Dubna, Russia}
\affiliation{GSI Helmholtzzentrum f\"ur Schwerionenforschung GmbH, Planckstra{\ss}e 1, 64291 Darmstadt, Germany}
\affiliation{Instituto de F\'{\i}sica,
	Universidad Aut\'onoma de San Luis Potos\'{\i}\\
	Av. Manuel Nava 6, San Luis Potos\'{\i}, S.L.P. 78290, M\'exico}

\author{D.~B.~Blaschke}
\email{blaschke@ift.uni.wroc.pl}
\affiliation{Bogoliubov  Laboratory of Theoretical Physics,
	Joint Institute for Nuclear Research,\\
	Joliot-Curie str. 6, 141980 Dubna,
	Russia}
\affiliation{Institute for Theoretical Physics,
	University of Wroc{\l}aw,
	Max Born Pl. 9,
	50-204 Wroc{\l}aw, Poland}
\affiliation{National Research Nuclear University (MEPhI),
	Kashirskoe Shosse 31,
	115409 Moscow, Russia}

\author{A.~G.~Grunfeld}
\affiliation{CONICET, Godoy Cruz 2290, (C1425FQB) Buenos Aires, Argentina}
\affiliation{Departamento de  F\'{\i}sica, Comisi\'on Nacional de Energ\'{\i}a At\'omica,
	Av. Libertador 8250, (1429) Buenos Aires, Argentina}

\author{V.~P.~Pagura}
\affiliation{Departamento de F\'{\i}sica Te\'orica and IFIC,
	Centro Mixto Universidad de Valencia-CSIC,
	E-46100 Burjassot (Valencia), Spain
}

\begin{abstract}
A class of hybrid compact star equations of state is investigated that joins by a Maxwell construction a low-density phase of hadronic matter, modeled by a relativistic meanfield approach with excluded nucleon volume, with a high-density phase of color superconducting two-flavor quark matter, described within a nonlocal covariant chiral quark model. We find the conditions on the vector meson coupling in the quark model under which a stable branch of hybrid compact stars occurs in the cases with and without diquark condensation. We show that these hybrid stars do not form a third family disconnected from the second family of ordinary neutron stars unless additional (de)confining effects are introduced with a density-dependent bag pressure. A suitably chosen density dependence of the vector meson coupling assures that at the same time the $2~$M$_\odot$ maximum mass constraint is fulfilled on the hybrid star branch.  A twofold interpolation method is realized which implements both, the density dependence of a confining bag pressure at the onset of the hadron-to-quark matter transition as well as the stiffening of quark matter at higher densities by a density-dependent vector meson coupling. For three parametrizations of this class of hybrid equation of state the properties of corresponding compact star sequences are presented, including mass twins of neutron and hybrid stars at 2.00, 1.39 and 1.20 $M_\odot$, respectively. The sensitivity of the hybrid equation of state and the corresponding compact star sequences to variations of the interpolation parameters at the 10\% level is investigated and it is found that the feature of third family solutions for compact stars is robust against such a variation. This advanced description of hybrid star matter allows to interpret GW170817 as a merger not only of two neutron stars but also of a neutron star with a hybrid star or of two hybrid stars.         
\end{abstract}
\maketitle

\section{Introduction}
Recently, in the context of the observation of pulsars with high masses of about $2~M_\odot$ by Demorest et al. \cite{Demorest:2010bx,Fonseca:2016tux,Arzoumanian:2017puf} and Antoniadis et al.~\cite{Antoniadis:2013pzd}, the question of the possible existence of a third family of compact stars was revived, because of its relation to strong phase transitions in dense matter \cite{Gerlach:1968zz}. The question was asked: could it be that stars at this high mass appear as mass twins \cite{Glendenning:1998ag} or almost mass twins like in the case above, where stars have about the same mass but may have very different radii, pointing to a very different structure of their interiors. In such a case, an explanation would be that the larger star would be an ordinary neutron star while the more compact one would exhibit a core composed of high-density matter, e.g., quark matter, with a large density jump at the border between inner core and outer core. The condition on the magnitude of the jump in energy density for the instability of neutron star configurations to occur is given by the Seidov relation \cite{Seidov:1971}.
In addition, the high-density matter has to be stiff enough to allow for a stable hybrid star branch, the so-called "third family" of compact stars.
The observational verification of the existence of mass twins would therefore imply very important lessons for the properties of compact star matter: the existence of a strong phase transition to a new form of high-density matter and the necessity that this matter obeys a sufficiently stiff equation of state (EoS). 
It is not only the case of high-mass twins which are interesting in this context. Also the possibility of mass twins in the range of typical neutron star masses around $1.4~M_\odot$ is very interesting! In such a case, the onset of the phase transition shall occur sufficiently early, at densities reached in the center of a typical-mass neutron star. It shall be soft enough at those densities to allow for a large enough density jump and stiffen quickly at increasing density to prevent gravitational collapse on the third family branch at least until the mass constraint of about $2~M_\odot$ is reached without violating the causality constraint. Namely that the speed of sound should not exceed the speed of light. 
If these constraints could be fulfilled, the corresponding EoS could provide a viable scenario for the recently observed \cite{TheLIGOScientific:2017qsa} binary neutron star merger event:
at least one of the two stars could be a hybrid star from the third family branch which is sufficiently compact to make the binary system fulfill the condition on the tidal deformabilities derived from the observation of the inspiral in the LIGO gravitational wave detector, see \cite{Paschalidis:2017qmb,Ayriyan:2017nby} for a recent discussion of this scenario.

Up to now, the third family case was investigated with very schematic EoS for the high-density phase, like the bag model \cite{Glendenning:1998ag}, the constant-speed-of-sound (CSS) model 
\cite{Alford:2013aca,Alvarez-Castillo:2013cxa,Blaschke:2013ana,Alford:2015dpa,Ranea-Sandoval:2015ldr,Alford:2017qgh,Christian:2017jni,Paschalidis:2017qmb}, 
the multi-polytrope model 
\cite{Read:2008iy,Alvarez-Castillo:2017qki,Paschalidis:2017qmb}, 
but also dynamical models for interacting quark matter like the Nambu-Jona-Lasinio (NJL) model with higher order quark interactions in the Dirac vector channel 
\cite{Benic:2014jia,Bejger:2016emu}, or a relativistic density-functional model
\cite{Kaltenborn:2017hus,Ayriyan:2017nby}.
The question arises whether the third family phenomenon could be obtained also for EoS derived from dynamical quark models which are closer to QCD in the sense that they take into account the running of the dynamical quark masses with 4-momentum and embody a (dynamical) confinement mechanism. Such models are provided by QCD Dyson-Schwinger equations in their generalization to finite temperature and chemical potential 
\cite{Roberts:2000aa} and have recently been applied to study hybrid compact stars, see, e.g., \cite{Klahn:2009mb,Chen:2016ran} and references therein. 
The dynamical model for the gluon propagator mediating the nonperturbative quark interactions is still quite schematic and more realistic ones including a matching with perturbative QCD behaviour at large momentum transfer 
make applications at high densities for compact star matter rather involved. 
Therefore, a separable ansatz for the nonlocal dynamical interactions has been suggested 
\cite{Schmidt:1994di,Bowler:1994ir} and was used for the description of hadron properties
in the QCD vacuum, see \cite{Burden:1996nh,Golli:1998rf,Radzhabov:2003hy,Dorokhov:2003kf} for early works.  
This ansatz allows to develop covariant nonlocal chiral quark models for low-energy QCD at finite temperatures and densities that share the running of the quark mass function 
\cite{Schmidt:1994di,GomezDumm:2006vz}
and the wave function renormalization of the quark propagator
\cite{Noguera:2008cm,Contrera:2010kz,Radzhabov:2010dd,Horvatic:2010md,Hell:2011ic}
with full QCD as probed in lattice QCD simulations
\cite{Parappilly:2005ei}.   
This approach allowed a description of the QCD phase diagram 
\cite{GomezDumm:2005hy,Hell:2009by,Contrera:2012wj,Contrera:2016rqj}
and has been also applied to the description of hybrid compact stars with quark matter cores \cite{Blaschke:2007ri}

For a recent discussion of the issues occuring in the description of phases of dense matter in compact stars under modern constraints see, e.g., Ref.~\cite{Blaschke:2018mqw}.
Models with the capability to address the occurrence of a third family of compact stars shall fulfill these conditions \cite{Benic:2014jia}
\begin{itemize}
\item stiff hadronic EoS (to obtain a phase transition in the observed mass range of compact stars)
\item stiff high-density EoS (to generate a stable hybrid star branch which should reach $2~M_\odot$),
\item sufficient density jump at the phase transition (to produce the instability as necessary condition for the existence of a third family branch).
\end{itemize}

Difficulties arise when one attempts to describe systematically a possible medium dependence of the interaction model that would go beyond the covariant formfactor ansatz and address Lorentz symmetry breaking effects \cite{Benic:2013eqa}.
Of particular importance for the description of cold degenerate QCD matter as in compact stars, a medium dependence of the vector meson interaction is a crucial effect as it directly related to the density dependence of the stiffness of QCD matter between the deconfinement transition and the perturbative QCD regime at asymptotically high densities.
In order to capture such nonperturbative medium effects on the basis of the nonlocal, covariant formfactor model it has been suggested in Ref.~\cite{Blaschke:2013rma}
to apply an interpolation technique similar in spirit to the one introduced for a flexible description of the deconfinement transition region in cold degenerate matter by Masuda et al.~\cite{Masuda:2012kf,Masuda:2012ed}
who followed earlier concepts of interpolation for the description of lattice QCD thermodynamics in the high-temperature region \cite{Asakawa:1995zu}.  
A recent discussion of this interpolation technique can be found in \cite{Baym:2017whm}
where also the necessity to capture softening effects from quark confinement has been pointed out.

In the present work we develop the interpolation technique for nonlocal covariant formfactor approaches further in order to fulfill the above requirements on models of compact star matter which would produce third family solutions for hybrid stars with quark matter cores that at the same time obey the constraints on their maximum mass to exceed the present observational bound of $2.01\pm 0.04$ M$_\odot$ \cite{Antoniadis:2013pzd} and on their compactness to be in accord with the bounds derived from the gravitational waves detected for the inspiral phase of the binary compact star merger GW170817 \cite{TheLIGOScientific:2017qsa}.
To this end we suggest here a twofold interpolation approach that captures the density dependence of both, the confining effects (the vanishing of a bag-like negative pressure of the nonperturbative QCD medium) and the stiffness (the increase of the vector meson 
coupling strength).

\section{Theoretical formalism}

In the present work we use a two-phase description in order to
account for the transition from nuclear to quark matter. 
In the following two subsections we explain the theoretical
approaches used to obtain the EoS for each of these phases.
Throughout this work we use natural units with $\hbar=c=k_B=G=1$.

\subsection{Nuclear matter equation of state}

For the nuclear matter phase, we use the well-established relativistic density-functional approach by Typel et al. \cite{Typel:2005ba} based on meson-exchange interactions and with the so-called "DD2" parametrization 
given in Ref. \cite{Typel:2009sy}. 
It describes the properties of nuclear matter at saturation density and below
very well, also in accordance with the chiral effective field theory approach
\cite{Kruger:2013kua}; see also Ref. \cite{Fischer:2013eka}. 
To improve the higher-density behavior, a generalized excluded volume effect is included according to Ref. \cite{Typel:2016srf}.
The DD2$\underline{~}$p40 EoS features this correction
by considering the available volume fraction $\Phi_N$ for the
motion of nucleons as density dependent in a Gaussian form
\begin{equation}
\Phi_N = \exp\left[-v|v|(n-n_0)^2/2\right],~{\rm for}~ n > n_0~,
\end{equation}
and $\Phi_N=1$ if $n\le n_0$.
Here, $v = 16\pi r_N^3/3$ is the van der Waals excluded volume
for a nucleon with a hard-core radius $r_N$ and $n_0=0.15$ fm$^3$ is the saturation density of infinite, symmetric nuclear matter. 
The index "p40" with the DD2 parametrization denotes a positive excluded volume parameter of $v=4$ fm$^3$.
This type of nuclear EoS has recently been extensively used in systematic studies of hybrid star models, see for instance \cite{Benic:2014jia,Alvarez-Castillo:2016oln,Kaltenborn:2017hus}.

In Fig.~\ref{fig:EoS} we show these two EoS in the form pressure $P$ vs. baryochemical potential $\mu$ (under conditions of $\beta-$equilibrium and charge neutrality), which is suitable to construct the phase transition to quark matter 
for the case of local charge conservation.
We show also the EoS "DD2F", for which the density-dependence of the meson-nucleon couplings at supersaturation densities is adjusted such that the EoS in the isospin-symmetric case fulfills the so-called "flow constraint" of Danielewicz et al. \cite{Danielewicz:2002pu}. 
We note that the stiffer the EoS, the flatter the curve in the P-$\mu$ diagram.
For comparison, we show the EoS of Akmal, Pandharipande and Ravenhall (APR), case "A18+$\delta v$+UIX*" from Ref~\cite{Akmal:1998cf}, which is soft at low densities and becomes very stiff at higher densities. 
All these hadronic EoS fulfill the constraint on the lower limit of the maximum mass of neutron stars from PSR J0348+0432 \cite{Antoniadis:2013pzd}, see Fig.~\ref{fig:M-R}.

\subsection{Quark matter equation of state}
\label{subsec:QMEoS}

For the description of the quark matter phase we consider
a nonlocal chiral quark model which includes scalar quark-antiquark
interaction, anti-triplet scalar diquark interactions and 
vector quark-antiquark interactions that was presented in detail in 
Ref. \cite{Blaschke:2007ri}. Let us start writing the
corresponding effective Euclidean action, that in the case of two light 
flavors reads \cite{Blaschke:2007ri}

\begin{eqnarray}
S_E &=& \int d^4 x \ \left\{ \bar \psi (x) \left(- i \rlap/\partial + m_c
\right) \psi (x) - \frac{G_S}{2} j^f_S(x) j^f_S(x) 
\right.\nonumber\\
&& \left. 
- \frac{H}{2}
\left[j^a_D(x)\right]^\dagger j^a_D(x) {-}
\frac{G_V}{2} j_V^{\mu}(x)\, j_V^{{\mu}}(x) \right\} \, .
\label{action}
\end{eqnarray}
Here $m_c$ is the current quark mass, that we consider to be equal for $u$
and $d$ quarks. The nonlocal currents $j_{S,D,V}(x)$ are given in terms of  
operators based on a separable approximation of the effective one gluon
exchange model (OGE) of QCD. These currents read
\begin{eqnarray}
j^f_S (x) &=& \int d^4 z \  g(z) \ \bar \psi(x+\frac{z}{2}) \ \Gamma_f\,
\psi(x-\frac{z}{2})\,,
\\
j^a_D (x) &=&  \int d^4 z \ g(z)\ \bar \psi_C(x+\frac{z}{2}) \ \Gamma_D \ \psi(x-\frac{z}{2}) 
\\
j^\mu_V (x) &=& \int d^4 z \ g(z)\ \bar \psi(x+\frac{z}{2})~ i\gamma^\mu
\ \psi(x-\frac{z}{2}). 
\label{cuOGE}
\end{eqnarray}
In the above equation we have used $\psi_C(x) = \gamma_2\gamma_4 \,\bar \psi^T(x)$,
$\Gamma_f=(\openone,i\gamma_5\vec\tau)$ and $\Gamma_D=i \gamma_5 \tau_2 \lambda_a$, 
while $\vec \tau$ and $\lambda_a$, with $a=2,5,7$, stand for Pauli and Gell-Mann 
matrices acting on flavor and color spaces, respectively. 
The function $g(z)$ in Eqs.~(\ref{cuOGE}) is a covariant formfactor characterizing 
the nonlocality of the effective quark interaction~\cite{GomezDumm:2005hy}.

The effective action in Eq.~(\ref{action}) might arise via Fierz
rearrangement from some underlying more fundamental interactions, and is
understood to be used ---at the mean field level--- in the Hartree
approximation. In general, the ratios of coupling constants $H/G_S$,
$G_V/G_S$ depend on such microscopic action. For example,
for OGE interactions in the vacuum Fierz transformation leads to 
$H/G_S =0.75$ and $G_V/G_S= 0.5$. 
However, since the precise form of the
microscopic interaction cannot be derived directly from
QCD, this value is subject to rather large theoretical uncertainties. 
In fact, thus far there is no strong phenomenological constraint
on the ratio $H/G_S$, except for the fact that values
larger than one, are quite unlikely to be realized in
QCD since they  might lead to color symmetry breaking
in the vacuum. 
We introduce $\eta=G_V/G_S$ for the dimensionless vector coupling 
strength and use it as a free parameter of the model rsponsible for the 
stiffness of the quark matter EoS at nonzero densities.
Details of the values used in the present work will be given below.

To proceed it is convenient to
perform a standard bosonization of the theory. Thus, we
introduce scalar, vector and diquark bosonic fields and integrate out 
the quark fields. In what follows we will work within the mean field approximation
(MFA), in which these bosonic fields are expanded
around their vacuum expectation values and the corresponding
fluctuations are neglected. The only nonvanishing mean field values 
in the scalar and vector sectors correspond to isospin zero fields, 
$\bar\sigma$ and $\bar\omega$ respectively.
Concerning the diquark mean field values, we will assume that
in the density region of interest only the 2SC phase might
be relevant, thus, we adopt the ansatz $\bar\Delta_5=\bar\Delta_7=0$,
$\bar\Delta_2=\bar\Delta$.

Next, we consider the Euclidean action at zero temperature and finite baryon
chemical potential $\mu_B$. For such purpose we introduce different chemical potentials
$\mu_{fc}$ for each flavor and color, then the corresponding mean field grand
canonical thermodynamic potential per unit volume can be written as

\begin{eqnarray}
\Omega^\mf  &=&   \frac{ \bar
\sigma^2 }{2 G_S} + \frac{ {\bar \Delta}^2}{2 H} 
- \frac{\bar \omega^2}{2 G_V} \nonumber\\
&&- \frac{1}{2} \int \frac{d^4 p}{(2\pi)^4} \ \ln
\mbox{det} \left[ \ S^{-1}(\bar \sigma ,\bar \Delta, \bar \omega,
\mu_{fc}) \right] \ , \label{mfaqmtp}
\end{eqnarray}
where we have introduced different chemical potentials
$\mu_{fc}$ for each flavor and color. The inverse propagator $S^{-1}$ is a $48 \times 48$ matrix in Dirac,
flavor, color and Nambu-Gorkov spaces. A detailed description of the model and the explicit 
expression for the thermodynamic potential after calculating the determinant of the inverse of the propagator
can be found in Ref.~\cite{Blaschke:2007ri}. 
Then, the mean field values $\bar \sigma$, $\bar \Delta$ and $\bar \omega$ can be obtained
by solving the coupled equations
\begin{eqnarray}
\frac{ d \Omega^\mf}{d\bar \Delta} \ = \ 0 \ , \ \ \
\frac{ d \Omega^\mf}{d\bar \sigma} \ = \ 0 \ , \ \ \
\frac{ d \Omega^\mf}{d\bar \omega} \ = \ 0 \ .
\label{gapeq}
\end{eqnarray}

In principle one has six different quark chemical potentials,
corresponding to quark flavors $u$ and $d$ and quark colors $r,g$ and $b$.
However, there is a residual color symmetry (between red and green
colors due to the the ansatz we considered) arising from the direction of $\bar\Delta$ in color space.
Moreover, if we require the system to be in chemical equilibrium, it can
be seen that chemical potentials are not independent from each other. In
general, it is shown that all $\mu_{fc}$ can be written in terms of three
independent quantities: the baryonic chemical potential $\mu$, a quark
electric chemical potential $\mu_{Q_q}$ and a color chemical potential
$\mu_8$. The corresponding relations read
\begin{eqnarray}
\mu_{ur} &=& \mu_{ug} = \frac{\mu}{3} + \frac23 \mu_{Q_q} + \frac13
\mu_8 \ ,\\
\mu_{ub} &=& \frac{\mu}{3} + \frac23 \mu_{Q_q} -
\frac23 \mu_8 \ , \\
\mu_{dr} &=& \mu_{dg} = \frac{\mu}{3} - \frac13 \mu_{Q_q} + \frac13
\mu_8 \ , \\
\mu_{db} &=& \frac{\mu}{3} - \frac13 \mu_{Q_q} - \frac23 \mu_8 \ \ .
\label{chemical}
\end{eqnarray}
The chemical potential $\mu_{Q_q}$, distinguishes between up
and down quarks, and the color chemical potential $\mu_8$ 
has to be introduced to ensure color neutrality.

As we are interested in describing the behaviour of quark matter in the core of
neutron stars we have to take into account
the presence of electrons and muons, in addition to quark matter. 
We treat leptons as a free relativistic Fermi gas and the corresponding
thermodynamic potential expression can be found in \cite{Blaschke:2007ri}. 
In addition, it is
necessary to take into account that quark matter has to be in beta
equilibrium with electrons and muons through the beta decay reactions
\begin{equation}
d\to u+l+\bar\nu_l\ ,~~~ { u+l\to d+\nu_l\ ,}
\end{equation}
for $l=e,\mu$. Thus, assuming that (anti)neutrinos escape from the
stellar core, we have an additional relation between fermion chemical
potentials, namely
\begin{equation}
\label{betaeq}
\mu_{dc} - \mu_{uc} = - \mu_{Q_q} = \mu_l
\end{equation}
for $c=r,g,b$, $\mu_e = \mu_\mu = \mu_l$.

Finally, in the core of neutron stars we also require the system to be
electric and color charge neutral, hence the number of independent
chemical potentials reduces further.
Indeed, $\mu_l$ and $\mu$ get fixed by the condition that charge and
color densities vanish,
\begin{eqnarray}
\label{charge}
\rho_{Q_{tot}} &=& \rho_{Q_q}- \sum_{l=e,\mu}\rho_l \nonumber \\ 
&=&\sum_{c=r,g,b} \left(\frac23 \rho_{uc} - \frac13 \rho_{dc} \right)
- \sum_{l=e,\mu}\rho_l = 0 \ , \\
\rho_8 & = & \frac{1}{\sqrt3} \sum_{f=u,d}
\left(\rho_{fr}+\rho_{fg}-2\rho_{fb} \right) \ = \ 0 \ ,
\label{dens}
\end{eqnarray}
where the expressions for the lepton densities $\rho_l$ and the quark
densities $\rho_{fc}$ can be found in the Appendix of \cite{Blaschke:2007ri}.
In summary, in the
case of neutron star quark matter, for each value of $\mu$ one can find
the values of $\bar \Delta$, $\bar \sigma$, $\mu_l$ and $\mu_8$
by solving the gap equations (\ref{gapeq}), supplemented by the constraints of Eq.~(\ref{betaeq}) 
for $\beta-$ equilibrium and Eqs. (\ref{charge}) and (\ref{dens}) for electric and color charge neutrality, resp. 
This allows to obtain the quark matter EoS in the relevant thermodynamic region as
\begin{equation}
\label{eq:P-mu}
P(\mu)=P(\mu;\eta(\mu),B(\mu))=-\Omega^\mf(\eta(\mu)) - B(\mu)~,
\end{equation}
where for later use we allow for the possibility of a bag pressure shift
stemming, e.g., from a medium dependence of the gluon sector, and both parameters $\eta$ and $B$ may depend on the chemical potential.

Such a generalized form of the nonlocal chiral quark model with {\it a priori} unknown $\mu-$dependences of the vector coupling strength and the bag pressure shift are suitable for Bayesian analyses of hybrid EoS to be constrained by sets of experimental data from heavy-ion collisions and/or compact star observations.

\subsection{Phase transition}
\label{ssec:PT}

In the present work, we shall use a simple Maxwell
construction for the phase transition between the 
EoS for nuclear matter and quark matter phases described  
above.
They shall separately fulfill charge-neutrality and 
beta equilibrium with electrons and muons.
The two distinct phases are then matched according to the 
Gibbs conditions for phase equilibrium by requiring 
that temperatures, chemical potentials and pressures of the 
two phases coincide at the phase transition
\begin{eqnarray}
T^H&=&T^Q=0,\\
\mu^H&=&\mu^Q=\mu_c,\\
P^H&=&P^Q=P_c~.
\end{eqnarray}
Technically, we plot the $T=0$ isotherms
of both phases in pressure over baryon
chemical potential and merge them at the crossing point
$P_c=P(\mu_c)$.
Outside the phase transition, the phase
with higher pressure (lower grand canonical potential)
is to be chosen as the physical one.

\begin{figure*}[!h]
	\includegraphics[width=0.9\textwidth]{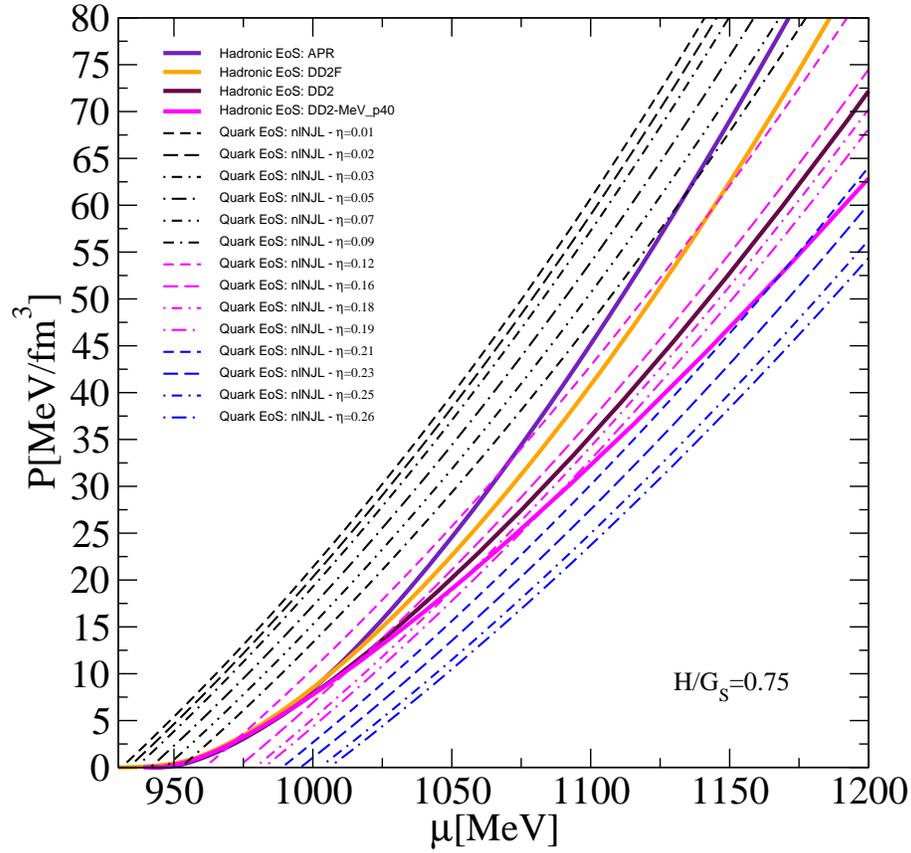}
	\caption{
		Quark matter EoS according to the nonlocal chiral quark model 
		with color superconductivity for different values of the dimensionless
		vector meson coupling strength parameter $\eta=0.01, 0.02, ..., 0.09$ (black thin lines), 
		$\eta=0.12, ..., 0.19$ (magenta thin lines), and $\eta=0.21, ..., 0.26$ (blue thin lines) compared with 
		four nuclear matter EoS ordered by increasing stiffness: APR (violet solid line), DD2F (orange solid line), DD2 (black solid line) and DD2\underline{ }p40 (magenta solid line). For a detailed discussion, see text.
		\label{fig:EoS}}
\end{figure*}

A more sophisticated construction of the phase transition considers
the occurrence of structures (so-called "pasta phases") with
an interplay of surface tension, Coulomb energy and
charge-screening effects. For a recent work, see e.g.
Ref.~\cite{Yasutake:2014oxa} and references therein.
It has been found that for typical values of the surface tension 
the pasta phase construction yields an EoS very similar to that of 
the Maxwell construction \cite{Voskresensky:2002hu}. 

\subsubsection{Color superconducting quark matter}

	In Fig.~\ref{fig:EoS} we show the results for the color superconducting, nonlocal chiral quark model EoS of the present work for different values of the dimensionless vector meson coupling parameter $\eta = 0.01, \dots, 0.26$.
Inspecting this figure we draw the following conclusions.
\begin{itemize}
	\item For the soft hadronic EoS (APR and DD2F)
no reasonable Maxwell construction of a phase transition is possible: 
for $\eta \ge 0.15$ quark and hadronic matter EoS do not cross, while for $\eta < 0.15$ the crossings do not describe a physically acceptable case - at low density quark matter would be thermodynamically favorable because of the larger pressure, while at higher densities a transition to hadronic matter would occur. 
\item For the DD2 EoS, we observe the "masquerade" situation: 
while for $\eta > 0.17$ there is no crossing of quark and hadronic EoS, and for $\eta \le 0.16$ the deconfinement transition would occur at unphysically low densities, for $\eta = 0.17$ at $\mu = 1050$ MeV the transition to a quark matter EoS occurs that is indistinguishable from the hadronic one. 
Then one would describe hybrid star matter that masquerades as neutron star
matter \cite{Alford:2004pf}.
\item For the stiff hadronic EoS DD2\underline{ }p40, one obtains deconfinement transitions which for $\eta \le 0.16$ occur at too low density,
but for $\eta \ge 0.17$ would result in acceptable hybrid star EoS with the specifics that for $\eta = 0.17$ a masquerade with the DD2 EoS is obtained.  
\end{itemize}
The corresponding solutions of the TOV equations are discussed in subsection \ref{ssec:M-R} below.

\subsubsection{Without color superconductivity}

\begin{figure*}
	\includegraphics[width=0.9\textwidth]{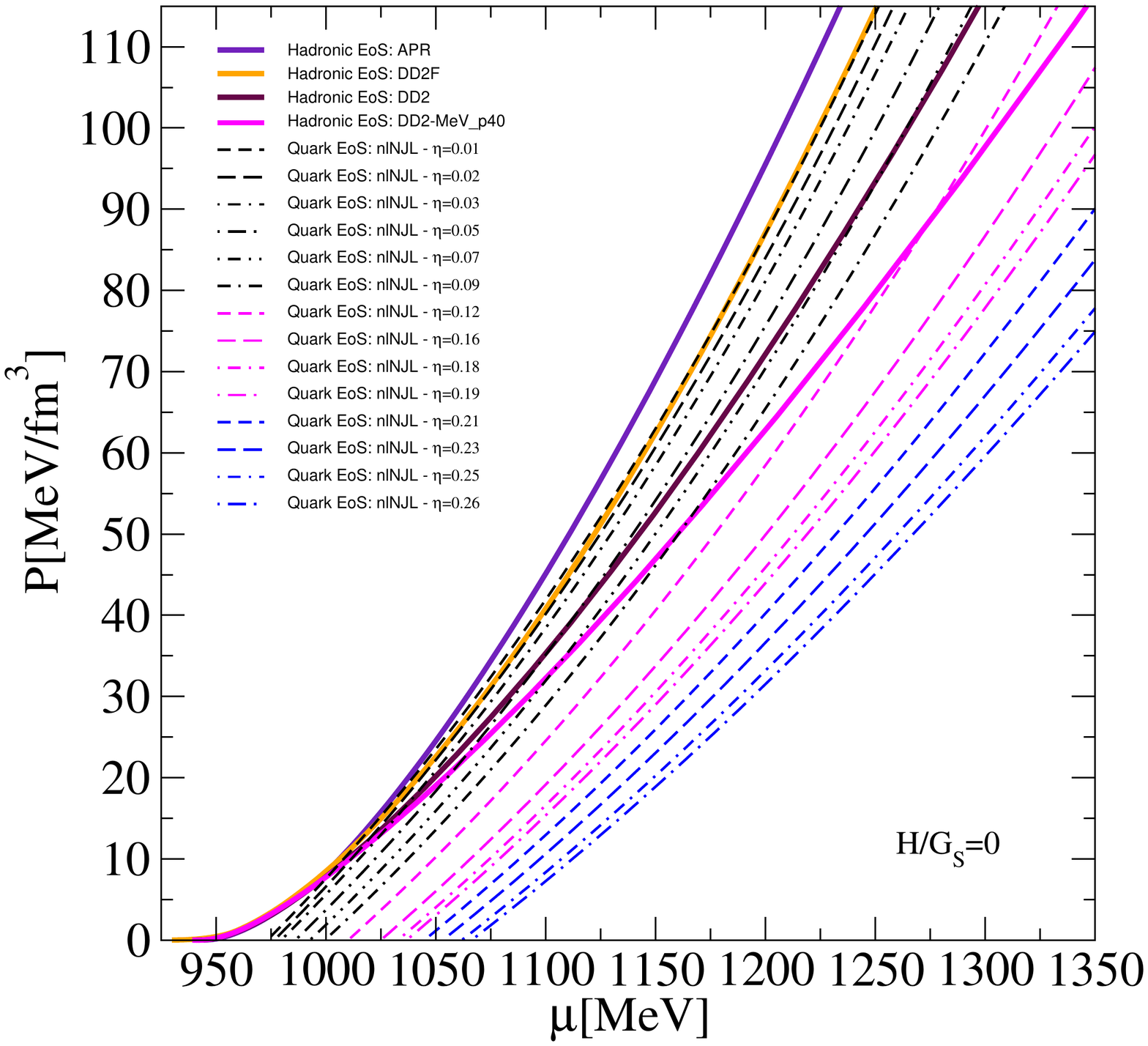}
	\caption{
		Quark matter EoS according to the nonlocal chiral quark model 
		without color superconductivity (diquark interaction switched off) for different values of the dimensionless
		vector meson coupling strength parameter $\eta=0.01, 0.02, ..., 0.09$ (black thin lines), 
		$\eta=0.12, ..., 0.19$ (magenta thin lines), and $\eta=0.21, ..., 0.26$ (blue thin lines) compared with 
		four nuclear matter EoS ordered by increasing stiffness: APR (violet solid line), DD2F (orange solid line), DD2 (black solid line) and DD2\underline{ }p40 (magenta solid line). For a detailed discussion, see text.
		\label{fig:EoS_nodiquark}}
\end{figure*}

Here we discuss the case when the effective coupling $H$ in the diquark interaction channel would be too small to result in diquark condensation, so that no color superconducting phase would occur.
Setting $H=0$ in Eq.~(\ref{action}), we obtain the set of quark matter EoS  
shown in Fig.~\ref{fig:EoS_nodiquark} together with the four hadronic EoS discussed also in Fig.~\ref{fig:EoS}.
When comparing both figures, we observe that for the soft hadronic EoS APR and DD2F no phase transition occurs for the whole set of vector couplings 
$\eta > 0.02$ and no reasonable phase transition construction exists for 
$\eta \le 0.02$.
For the stiffer hadronic EoS DD2 and DD2\underline{ }p40 the deconfinement transition is obtained by Maxwell construction with all values of the vector coupling $\eta$, albeit at rather low densities for  values $\eta \le 0.03$.
The corresponding solutions of the TOV equations are also discussed in subsection \ref{ssec:M-R} below.

\subsection{TOV equations, moment of inertia and tidal deformability}

In this subsection, we present the set of equations that are to be solved for obtaining the numerical results on compact star structure and global properties shown in the next section, once the EoS is given.

\subsubsection{TOV equations}

In order to compute the internal energy density distribution of compact stars and thus derive the mass-radius relation we utilize the Tolman--Oppenheimer--Volkoff (TOV) equations for a static and spherical star in the framework of general relativity~\cite{Tolman:1939jz,Oppenheimer:1939ne}:
\bea
\label{eq:TOVa}
\frac{dP( r)}{dr}&=& 
-\frac{\left(\varepsilon( r)+P( r)\right)
\left(m( r)+ 4\pi r^3 P( r)\right)}{r\left(r- 2m( r)\right)},\\
\frac{dm( r)}{dr}&=& 4\pi r^2 \varepsilon( r).
\label{eq:TOVb}
\eea
 By considering that $P(r=R)=0$, $M=m(r=R)$ and $ P_c= P(r=0)$  we have the necessary boundary and initial conditions for a relativistic star with mass $M$ and radius $R$, respectively. 

Once a central density $\epsilon_c$ is chosen, the TOV solutions provide internal density profiles of a star.  By increasing $\epsilon_c$ (or equivalently $P_c$)  for each star, the whole sequence up to the maximum mass, the mass-radius relation can be computed. A similar relation can be derived for the baryon number $N_B( r)$, enclosed in a distance $r$ from the center of the mass distribution,
\bea
\frac{d N_B( r)}{dr}&=& 4\pi r^2 \left(1-\frac{2m( r)}{r}\right)^{-1/2}n( r)~.
\eea
Integrating this equation one ontains the baryon number $N_B=N_B(R)$ of the star, from which its baryonic mass is obtained as $M_B=m_N N_B$, where $m_N$ is the nucleon mass. 

\subsubsection{Moment of inertia}

Another important stellar quantity is the moment of inertia. We compute the relativistic moment of inertia based on the approach presented in~\cite{Ravenhall:1994}
\begin{eqnarray}
 I &\simeq& \frac{J}{1+2J/R^{3}}~,\\
 J &=&\frac{8\pi}{3}\int_{0}^{R}  dr r^{4}\frac{\varepsilon(r)+P(r)}{1-2m(r)/r}. 
\end{eqnarray}
For a detailed discussion of the moment of inertia in the slow-rotation approximation, and for the hybrid star case see, e.g., \cite{Chubarian:1999yn,Zdunik:2005kh,Bejger:2016emu}, and references therein. 

\subsubsection{Tidal Love number and deformability parameter $\Lambda$}

In this section we briefly describe how to compute the tidal deformability (TD) of a compact star, based on the results of ~\cite{Hinderer:2007mb,Damour:2009vw,Binnington:2009bb,Yagi:2013awa,Hinderer:2009ca}.
We start by considering the dimensionless tidal deformability  parameter $\Lambda=\lambda/M^{5}$ which is computed for small tidal deformabilities.  Here $\lambda$ is the stellar TD and $M$ is the stellar gravitational mass  as introduced before. $\lambda$ is related to the so called love number 
\begin{equation} 
k_2 = \frac{3}{2} \lambda R^{-5}.
\label{k2def}
\end{equation}
The TD can be thought of a  modification of the space-time metric by a linear $l=2$ perturbation onto the spherically symmetric star, 
\begin{eqnarray}
ds^2 &=& - e^{2\Phi(r)} \left[1 + H(r) Y_{20}(\theta,\varphi)\right]dt^2
\nonumber\\
& & + e^{2\Lambda(r)} \left[1 - H(r) Y_{20}(\theta,\varphi)\right]dr^2
\nonumber \\
& & + r^2 \left[1-K(r) Y_{20}(\theta,\varphi)\right] \left( d\theta^2+ \sin^2\theta
d\varphi^2 \right),
\nonumber\\
& &
\end{eqnarray}
where $K'(r)=H'(r)+2 H(r) \Phi'(r)$, primes denoting derivatives with respect to $r$. 
The functions
$H(r)$, $\beta(r) = dH/dr$ obey
\begin{eqnarray}
\frac{dH}{dr}&=& \beta\\
\frac{d\beta}{dr}&=&2 \left(1 - 2\frac{m(r)}{r}\right)^{-1} \nonumber\\
&& H\left\{-2\pi
  \left[5\varepsilon(r)+9 P(r)+f(\varepsilon(r)+P(r))\right]\phantom{\frac{3}{r^2}} \right. \nonumber\\
&& \left.+\frac{3}{r^2}+2\left(1 - 2\frac{m(r)}{r}\right)^{-1}
  \left(\frac{m(r)}{r^2}+4\pi r P(r)\right)^2\right\}\nonumber\\
&&+\frac{2\beta}{r}\left(1 -
  2\frac{m(r)}{r}\right)^{-1}\nonumber\\
&&  \left\{-1+\frac{m(r)}{r}+2\pi r^2
  (\varepsilon(r)-P(r))\right\}~,
  \end{eqnarray}
where $f =d\epsilon/dp$ is the equation of state. 
The above equations must be solved simultaneously with the TOV equations. 
The system is to be integrated outward starting near the center using the expansions $H(r)=a_0 r^2$ and $\beta(r)=2a_0r$ as $r \to 0$. 
$a_0$ is a constant that determines how much the star is deformed and turns out to be 
arbitrary since it cancels in the expression for the Love number. With the definition of
\begin{equation}
y = \frac{ R\, \beta(R)} {H(R)},
\end{equation}
the $l=2$ Love number is found as
\begin{eqnarray}
k_2 &=& \frac{8C^5}{5}(1-2C)^2[2+2C(y-1)-y]\nonumber\\
      & & \times\bigg\{2C[6-3y+3C(5y-8)]\nonumber\\
      & &+4C^3[13-11y+C(3y-2)+2C^2(1+y)]\nonumber\\
      & &+3(1-2C)^2[2-y+2C(y-1)] \ln(1-2C)\bigg\}^{-1},\nonumber\\
\label{eq:k2}
\end{eqnarray}
where $C=M/R$ is the compactness of the star.

\begin{figure*}
	\includegraphics[width=0.9\textwidth]{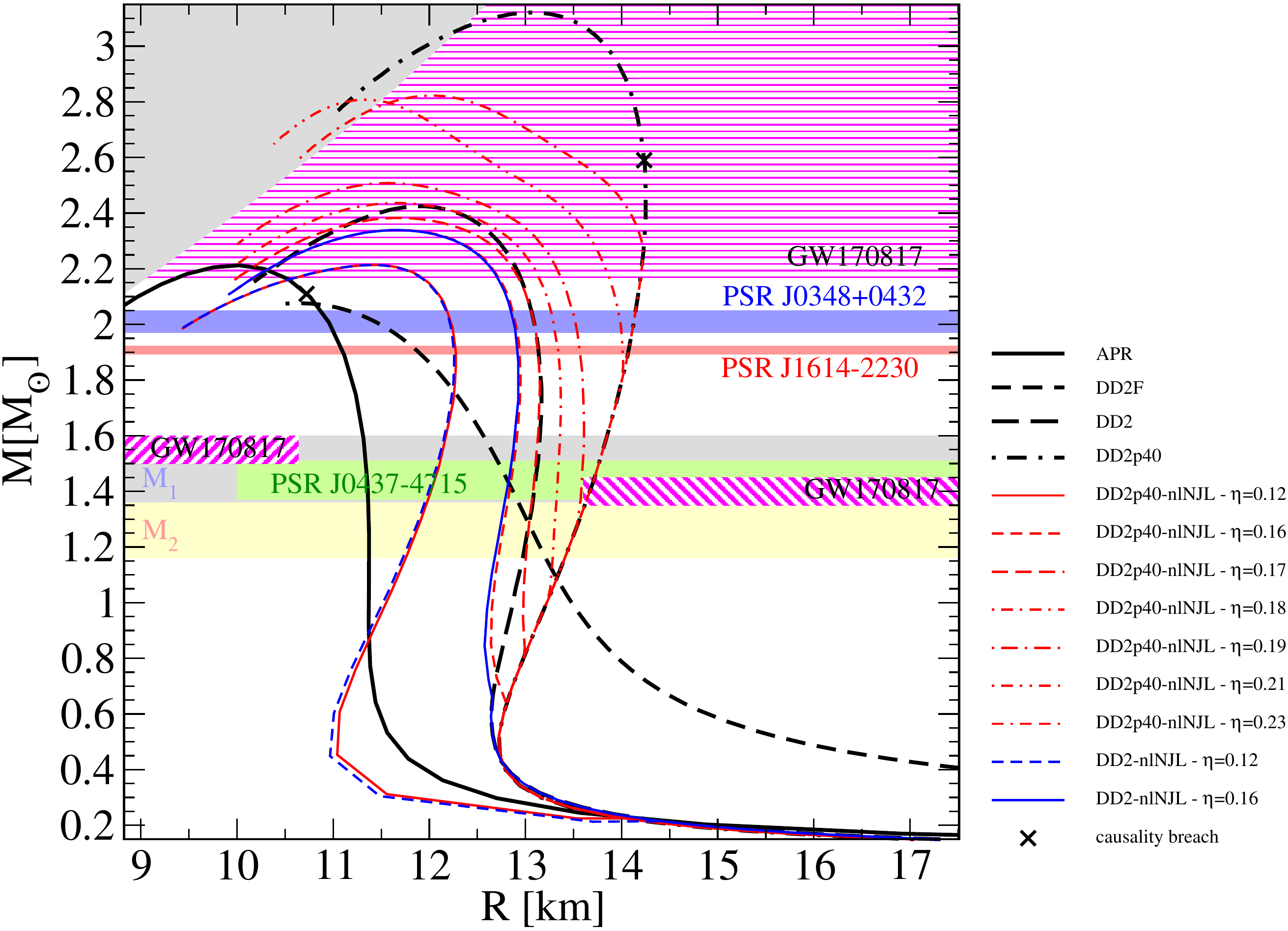}
	\caption{
		Mass-radius sequences for the class of hybrid EoS emerging from Maxwell constructions of standard nuclear EoS and the color superconducting nonlocal chiral quark model EoS of the present work, as discussed in subsect.~\ref{ssec:PT} and in Fig.~\ref{fig:EoS}. 
		\label{fig:M-R}}
\end{figure*}

\begin{figure*}
	\includegraphics[width=0.9\textwidth]{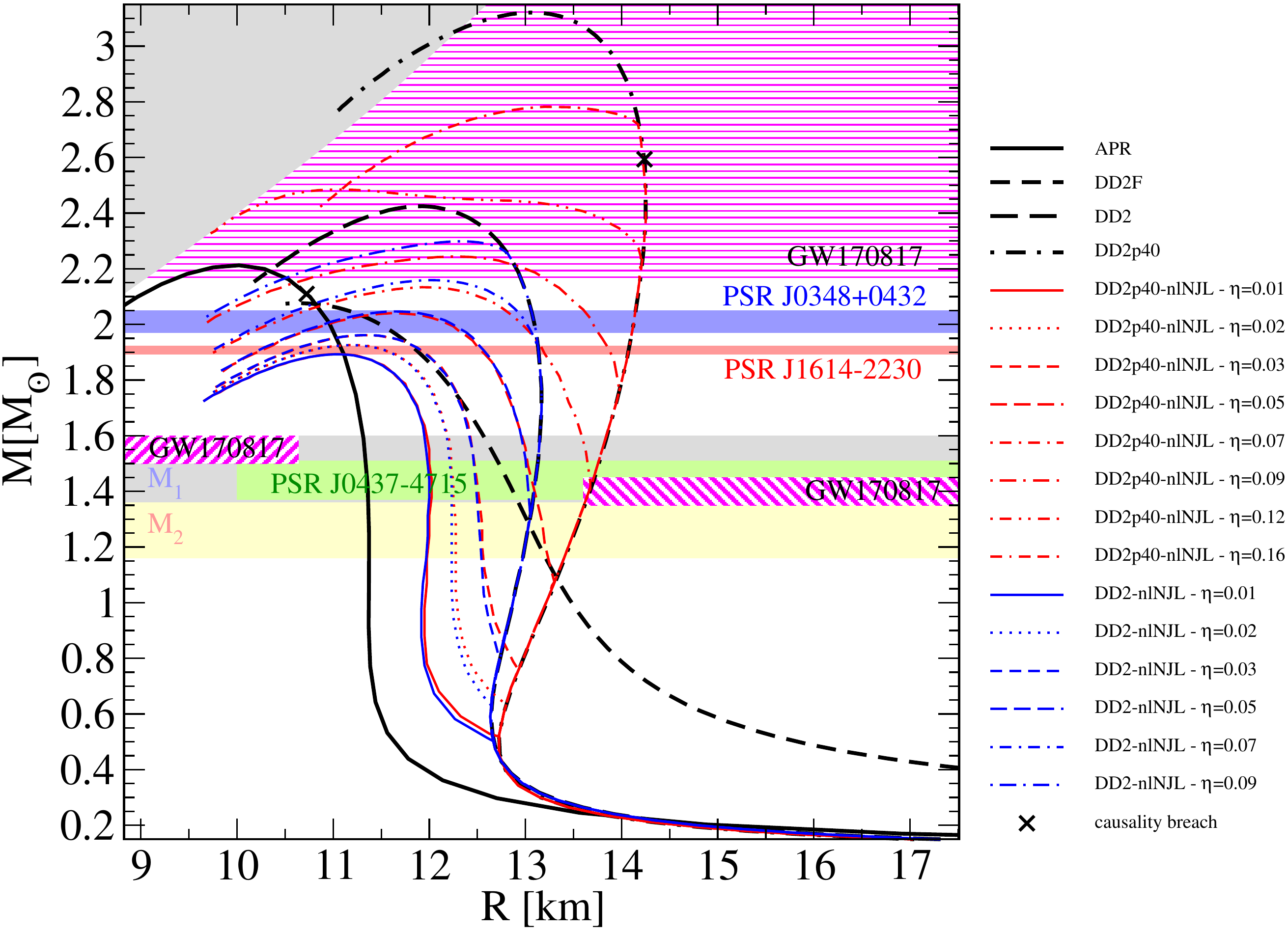}
	\caption{
		Mass-radius sequences for the class of hybrid EoS emerging from Maxwell constructions of standard nuclear EoS and the nonlocal chiral quark model EoS of the present work, where the color superconducting diquark channel is switched off, as discussed in subsect.~\ref{ssec:PT} and in Fig.~\ref{fig:EoS_nodiquark}. 
		\label{fig:M-R_nodiquark}}
\end{figure*}

\subsection{$M-R$ relations for hybrid star EoS with constant coupling strengths}
\label{ssec:M-R}
In Fig.~\ref{fig:M-R} we show the results for the compact star sequences in the $M-R$ diagram that result from the integration of the TOV equations 
(\ref{eq:TOVa}) and (\ref{eq:TOVb}) with the EoS of Fig.~\ref{fig:EoS}.
In this figure we also show by the blue band the lower limit 
$M_{\rm max}^{\rm low}=2.01\pm 0.04$ M$_\odot$
\cite{Antoniadis:2013pzd} for the maximum mass of a sequence that is fulfilled for all EoS.
New astrophysical constraints are derived from the recent observation of the compact star merger GW170817 \cite{TheLIGOScientific:2017qsa} as a minimal radius at $M=1.6~$M$_\odot$ of $R_{1.6}\ge 10.7$ km \cite{Bauswein:2017vtn}, 
a maximal radius at $M=1.4~$M$_\odot$ of $R_{1.4}\le 13.6$ km \cite{Annala:2017llu} and a possible upper limit on the maximum mass of nonrotating star configurations of $M_{\rm max}^{\rm up}=2.16\pm 0.04$ M$_\odot$ \cite{Rezzolla:2017aly}.
While the radius constraints are fulfilled by all EoS ($R_{1.4}$ only marginally by DD2\underline{ }p40), the new upper limit on the maximum mass is violated by all EoS except DD2F. 
Moreover, there limiting masses above which the EoS in the center of the star goes superluminal (indicated by crosses in Fig.~\ref{fig:M-R}).

The main conclusion we want to draw in the context of the present work is that the hybrid star sequences shown as red and blue thin lines with different line styles are all connected to their hadronic parent sequences. 
No third families of hybrid stars are possible with the nonlocal chiral quark model and constant coupling strengths. Therefore, there are also no mass twin stars.   

The same conclusion can be drawn for the case when color superconductivity in quark matter is switched off, see Fig.~\ref{fig:M-R_nodiquark}.
However, here there are hybrid star sequences possible that fulfill also the new constraint of the upper limit for the maximum mass, namely for the vector coupling strengths $0.05<\eta<0.07$ with either the DD2 or the 
DD2\underline{ }p40 hadronic EoS.

\section{Reconstruction of a target EoS}

In the previous section we have seen that for all parametrizations of the nonlocal chiral quark model with constant coupling strengths, one does not obtain a third family of hybrid stars (and thus twin stars). A reasonable Maxwell construction is obtained only with stiffer hadronic EoS like DD2 and DD2\underline{ }p40 and results then in hybrid star branches that are connected to the hadronic sequence, see Fig.~\ref{fig:M-R}.

In order to obtain twin stars and disconnected hybrid star branches using the nonlocal chiral model, we will generalize it by allowing a medium dependence of the vector and scalar meson coupling strength. 
This would be equivalent to a formulation with medium dependent vector coupling and pressure shift (bag pressure) for a constant scalar coupling. 
As a strategy to fix these {\it a priori} unknown density dependences we suggest here to adjust them such as to reproduce a target EoS with desired properties. The EoS of our choice is that of the recently developed relativistic string-flip model (SFM) \cite{Kaltenborn:2017hus} which produces stable disconnected hybrid star branches with an onset mass of the phase transition that depends on the model parameter  $\alpha$ that regulates the screening of the string tension. We will reconstruct the SFM EoS for the cases of $\alpha=0.2$ and $0.3$.

A third set is defined here by lowering the parameter $\mu_<$ for the onset of the deconfinement transition which results in a still lower mass $M_c=1.20~$M$_\odot$ for the onset of the neutron star instability against formation of the hybrid star sequence.

\subsection{Reconstruction by interpolation}

When asymptotic forms of the EoS are known, e.g., in the low-density regime of nuclear matter and in the high-density regime of quark matter, but they cannot readily be trusted in the intermediate range of densities where a quark-hadron phase transition is expected, it has been suggested in Ref.~\cite{Masuda:2012kf} to use an interpolation technique (a corrected version has been given in \cite{Masuda:2012ed}).
These works were inspired by the earlier developed interpolation technique 
\cite{Asakawa:1995zu} for the EoS in the vicinity of the crossover transition at finite temperatures and vanishing baryon densities. 

Such an interpolation technique has been used subsequently for the nlNJL EoS  $(P(\mu;\eta))$ which depends on an a priori unknown value of the vector coupling strength parameter $\eta$. 
Since this parameter may actually vary as a function of the chemical potential, it has been suggested in \cite{Blaschke:2013rma} to use the interpolation technique in order to model the EoS with a varying vector coupling strength parameter within certain limits and in a certain range of chemical potentials.

In the present work we want to generalize this approach by using it for two purposes:
\begin{enumerate}
	\item to model the unknown density dependence of the confining mechanism by interpolating a bag pressure contribution between zero and a finite value $B$ at low densities in the vicinity of the hadron-to-quark matter transition, and
	\item to model a density dependent stiffening of the quark matter EoS at high 
	density by interpolating between EoS for two values of the vector coupling strength, 
	$\eta_<$ and $\eta_>$. 
\end{enumerate}
The resulting doubly interpolated quark matter EoS reads
\begin{eqnarray}
\label{eq:twofold}
P(\mu) &=& [f_<(\mu) P(\mu;\eta_<,B) + f_>(\mu) P(\mu;\eta_<,0)] f_{\ll}(\mu)
\nonumber\\
&& +f_{\gg}(\mu)P(\mu;\eta_>,0)~,
\end{eqnarray}
where we have introduced two smooth switch-off functions, one that changes from one to zero at a lower chemical potential $\mu_<$ with a width $\Gamma_<$, 
\begin{equation}
\label{eq:f<}
f_{<}(\mu)=\frac{1}{2}\left[1-\tanh\left(\frac{\mu-\mu_<}{\Gamma_<}\right)\right],
\end{equation}
and one that switches off at a higher chemical potential $\mu_\ll$ with a width $\Gamma_\ll$, 
\begin{equation}
\label{eq:f<<}
f_{\ll}(\mu)=\frac{1}{2}\left[1-\tanh\left(\frac{\mu-\mu_\ll}{\Gamma_\ll}\right)\right],
\end{equation}
whereby the corresponding switch-on functions are the complementary ones,
\begin{equation}
f_{>}(\mu) = 1 - f_<(\mu)~,~~f_{\gg}(\mu) = 1 - f_{\ll}(\mu).
\end{equation}

The input EoS differ in their $\eta$-values, we have taken for those values at low ($<$) and high ($>$) chemical potentials 
$\eta_< $ and $\eta_> $. Moreover, $B$ represents a bag constant introduced to enforce confinement effects in the low chemical potential quark EoS. 

More detailed discussions about the adoption of a bag constant (or bag function) in addition to the chiral quark model pressure see, e.g., 
Refs.~\cite{Pagliara:2007ph,Blaschke:2010vj,Bonanno:2011ch,Klahn:2015mfa}.
Values of $B=10 - 50$ MeV/fm$^3$ are used in these references and will be taken as a guidance for adjusting this parameter in the present work. For a density-dependent bag constant in hybrid compact star physics applications, see also Ref.~\cite{Maieron:2004af}.

Note that for $B=0$ one would not obtain third family solutions within the nonlocal NJL model approach. 

\subsection{Equivalence of interpolation and density-dependent parameters}
\label{ssec:equiv}

\subsubsection{Bag pressure shift}
\label{sssec:bag}

Here we prove that the interpolation between two EoS that differ by a constant shift of the pressure can be rewritten as a pressure with $\mu-$dependent shift that is defined by the switch function.

\begin{eqnarray}
P(\mu) &=& P(\mu;\eta,B)f_<(\mu) + P(\mu;\eta,0)f_>(\mu)\nonumber\\
&=&P(\mu;\eta,0)[f_<(\mu)+f_>(\mu)] - B f_<(\mu) \nonumber\\
&=& P(\mu;\eta,B(\mu)),
\end{eqnarray}  
where 
\begin{equation}
\label{eq:B-mu}
B(\mu)=Bf_<(\mu)
\end{equation} 
is the $\mu-$dependent bag pressure.
Numerical results for the parametrisations defined in set 1 - set 3 are shown in Fig.~\ref{fig:B-mu}.

\subsubsection{Vector meson coupling}
\label{sssec:vector}

Here we prove that for small changes in the vector meson coupling $\eta$, when a first order Taylor expansion will be a sufficiently accurate approximation, the result of the interpolation may again be reinterpreted as a $\mu-$dependence of the vector meson coupling, defined by the switch functions $f_<(\mu)$ and $f_>(\mu)$.
\begin{eqnarray}
P(\mu) &=& P(\mu;\eta_<,B)f_\ll(\mu) + P(\mu;\eta_>,B)f_\gg(\mu)\nonumber\\
&=&P(\mu;\eta_<,B)[f_\ll(\mu)+f_\gg(\mu)] \nonumber\\
&&+(\eta_>-\eta_<) f_\gg(\mu) \frac{dP(\mu;\eta,B)}{d\eta}\bigg|_{\eta=\eta_<}\nonumber\\
&=& P(\mu;\eta_<,B)\nonumber\\
&&+[\eta_> f_\gg(\mu)+ \eta_< f_\ll(\mu)-\eta_<] \frac{dP(\mu;\eta,B)}{d\eta}\bigg|_{\eta=\eta_<},\nonumber\\
&=& P(\mu;\eta(\mu),B)~,
\end{eqnarray}  
where 
\begin{equation}
\label{eq:eta-mu}
\eta(\mu)=\eta_> f_\gg(\mu)+ \eta_< f_\ll(\mu)
\end{equation} 
is the medium-dependent vector meson coupling strength resulting from the pressure interpolation for the two values $\eta_<$ and $\eta_>$ with the switch functions $f_\ll(\mu)$ and $f_\gg(\mu)$. 

{
\subsubsection{Combined effect on $B(\mu)$ and $\eta(\mu)$ in the twofold interpolation}
\label{sssec:combine}

We want to express the twofold interpolated EoS of Eq.~(\ref{eq:twofold}) in terms of the generalized nonlocal chiral quark model (\ref{eq:P-mu}) with $\mu$-dependent parameters. To this end, starting from (\ref{eq:twofold}) and applying the relations derived in subsubsections \ref{sssec:bag}
and \ref{sssec:vector}, we get
\begin{eqnarray}
\label{eq:twofold2}
P(\mu) &=& [f_<(\mu) P(\mu;\eta_<,B) + f_>(\mu) P(\mu;\eta_<,0)] f_{\ll}(\mu)
\nonumber\\
&& +f_{\gg}(\mu)P(\mu;\eta_>,0)~,
\nonumber\\
&=&[P(\mu;\eta_<,0) - B f_<(\mu)] f_\ll (\mu) + P(\mu;\eta_>,0) f_\gg (\mu)
\nonumber\\
&=& P(\mu;\eta(\mu),0) - B(\mu)~,
\end{eqnarray}
where $\eta(\mu)$ is given by Eq.~(\ref{eq:eta-mu}) and the bag function for the twofold interpolation is
\begin{equation}
\label{eq:bag}
B(\mu)=Bf_<(\mu)f_\ll(\mu)~.
\end{equation} 
We note that this bag function differs from that of the onefold interpolation (\ref{eq:B-mu}), in particular when the region of the switches defined by the functions $f_<(\mu)$ and $f_\ll(\mu)$ do overlap.
}

\section{Numerical results and discussion}

In the present work we consider the case of a Gaussian form factor function $g(z)$
which translates to a Gaussian regulator function 
$g(p)=\exp (-p^2/p_0^2)$ in Euclidean 4-momentum space.
The fixed parameters of the quark model are $m_c=5.4869$ MeV, $p_0=782.16$ MeV, 
$G_S p_0^2 = 19.804$ and $H/G_S=0.75$.
{
For a better overview on the results reported below we introduce subsections.

\subsection{Twofold interpolation and generalized nonlocal chiral quark model}
}
The dimensionless vector coupling $\eta$ and the bag pressure $B$ are the two parameters that are used for calculating the three quark matter EoS that enter the twofold interpolation (\ref{eq:twofold}) which itself is determined by the four parameters $\mu_<$, $\Gamma_<$ and $\mu_\ll$, $\Gamma_\ll$ for the switch functions $f_<(\mu)$ and $f_\ll (\mu)$ given in Eqs.~(\ref{eq:f<}) and (\ref{eq:f<<}), respectively.
In table~\ref{tab:param} we give the values for three sets of these seven parameters which will be used in the numerical calculations of this work.
{
Set 1 and set 2 are defined such that they reproduce the two target EoS of Ref.~\cite{Kaltenborn:2017hus} that produce third family branches with twin star configurations in the M-R diagram at high masses ($\alpha=0.2$) and at low masses ($\alpha=0.3$) in their Fig.~15 for the corresponding EoS shown in Fig.~13 of that work. 
Our set 3 is a new result, representing an example from the class of hybrid star EoS that can be generated with the generalization of the nonlocal chiral quark model developed in this work.
This set is adjusted such as to have a still lower mass onset of the hybrid star branch than set 2 so that the binary compact star merger GW170817 could be explained as a merger of two hybrid stars with quark matter cores.
}

\begin{table}[htb]
	\begin{tabular}{|l|l|c|c|c|}
		\hline
		no&parameter& set 1 & set 2 & set 3\\
		\hline \hline
		1&$\mu_{<}$ [MeV] &  1600 & 1150& 1090\\
		2&$\Gamma_{< }$ [MeV] & 270 & 170& 170\\
		3&$\mu_{\ll}$ [MeV] & 1500 & 1700& 1700\\
		4&$\Gamma_{\ll}$ [MeV] & 300 & 300& 300\\
		5&$\eta_<$ & 0.09 & 0.05 & 0.05\\
		6&$\eta_>$ & 0.12 & 0.12 & 0.12\\
		7&$B$ [MeV/fm$^3$] & 35 & 35& 35\\
		\hline
	\end{tabular}
	\caption{Parameter sets 1 - 3 for the interpolated nonlocal NJL model. 
		\label{tab:param}
}  
\end{table}
\begin{table}[htb]
	\begin{tabular}{|l|c|c|c|}
		\hline
		& set 1 & set 2 & set 3\\
		\hline \hline
		$\mu_c$ [MeV] & 1214.06 & 1102.99& 1077.29 \\
		$p_c$[MeV/fm$^{3}$] & 67.5285 & 33.1236 & 26.0674 \\ 
		$\epsilon_c$ [MeV/fm$^{3}$] & 338.761 & 277.373 & 260.396 \\
		$n_c$ [fm$^{-3}]$ & 0.334568 &  0.281437
		 & 0.265871\\
		\hline
		$M_{c}$ [$M_\odot$] &2.00 & 1.39& 1.20\\
		$M_{\rm min}$ [$M_\odot$]&1.987 & 1.349& 1.166 \\
		$M_{\rm max}$ [$M_\odot$]&2.058 & 2.041 & 2.058 \\
		\hline
	\end{tabular}
	\caption{
	Results for a hybrid EoS with a first-order phase transition from hadronic matter described by the DD2$\underline{~}$p40 EoS 	\cite{Typel:2016srf}
		to quark matter described by the interpolated nonlocal NJL model (\ref{eq:twofold2}) obtained by a Maxwell construction for the three parametrizations of sets 1 - 3 given in Tab.~\ref{tab:param}: the critical chemical potential 
		$\mu_c$, the critical pressure $p_c$, and the values of energy density and baryon number density corresponding to the onset of the first-order phase transition, $\varepsilon_c$ and $n_c$, resp.
		Upon solving the TOV equations with the hybrid EoS, the mass-radius relation for compact stars is obtained which reveals a maximum mass $M_{\rm max}$ and a minimum mass $M_{\rm min}$ of the hybrid star branch as well as a mass $M_c$ at the onset of the phase transition in the center of the compact star.
		\label{tab:results}
}  
\end{table}

\begin{figure}[!th]
	\includegraphics[width=0.6\textwidth]{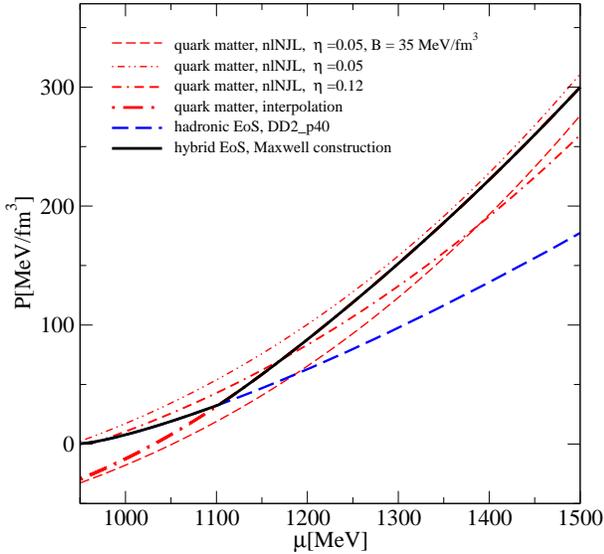}
	\vspace{-10mm}
	\caption{Hybrid star EoS (black solid line) obtained by a Maxwell construction between the quark matter EoS for set 2 (red dash-dotted line) and the density-dependent relativistic meanfield EoS DD2$\underline{~}$p40 for nuclear matter in $\beta-$equilibrium with electrons and muons.
		The doubly interpolated quark matter EoS is based on three parametrizations of the nonlocal NJL model: a soft (low vector coupling $\eta$) one with confinement ($B\neq 0$) at low densities (red dashed line), a soft one without confinement at intermediate densities (red dash-double-dotted line) and a stiff one (high $\eta$) at high densities (red double-dash-dotted line). The parameters of the switching functions are given in table \ref{tab:param}.
		\label{fig:1a}}
\end{figure}
  
In Fig.~\ref{fig:1a} we show for the example of parameter set 2 of Tab.~\ref{tab:param} the three quark matter EoS (in red thin lines) generated by the nonlocal chiral quark model that are used in the twofold interpolation procedure of Eq.~(\ref{eq:twofold}) to obtain the resulting quark matter EoS shown by the thick red dash-dotted line 
{ 
which can be reconstructed as an EoS of the generalized nonlocal chiral quark model (\ref{eq:twofold}) with the $\mu$-dependent bag function $B(\mu)$ from Eq.~(\ref{eq:bag}) and vector coupling $\eta(\mu)$ from Eq.~(\ref{eq:eta-mu}).
}

\begin{figure}[!th]
	\includegraphics[width=0.6\textwidth]{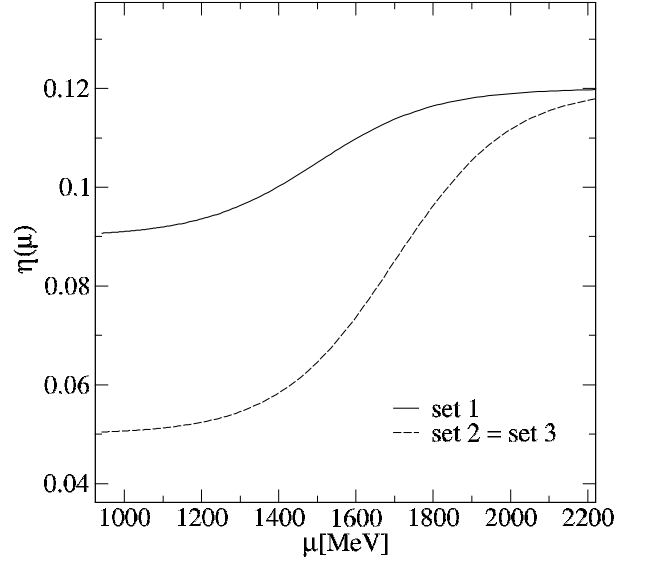}
	\vspace{-10mm}
	\caption{Medium dependence of the dimensionless vector meson coupling $\eta(\mu)$ as defined by the parameters of the interpolation procedure given in table \ref{tab:param} for sets 1, 2 and 3.		
		\label{fig:eta-mu}}
\end{figure}

\begin{figure}[!th]
	\includegraphics[width=0.6\textwidth]{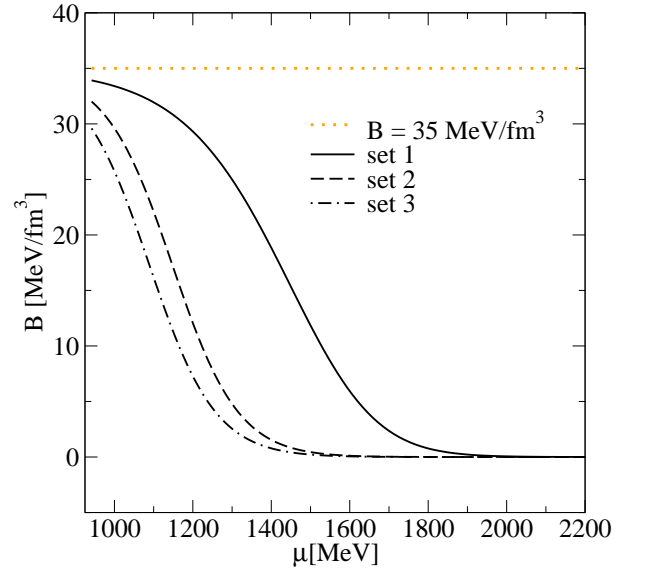}
	\vspace{-10mm}
	\caption{Medium dependence of the Bag pressure $B(\mu)$ as defined by the parameters of the interpolation procedure given in table \ref{tab:param} for sets 1, 2 and 3.		
	\label{fig:B-mu}}
\end{figure}

{
The explicit formulas for this reconstruction are a main result of this work, given in subsection
\ref{ssec:equiv}.  
This procedure describes a stiffening of the EoS at high densities due to an increase in the vector coupling strength $\eta(\mu)$ shown in 
Fig.~\ref{fig:eta-mu} and a softening due to onset of confinement at low densities as described by the bag pressure function $B(\mu)$ shown in
Fig. \ref{fig:B-mu}.

A Maxwell construction is performed with the hadronic EoS DD2$\underline{~}$p40 from \cite{Typel:2016srf} 
}
shown by the blue dashed line which results in the black solid line for the quark-hadron hybrid EoS.
 
The three parameter sets  given in table \ref{tab:param} are adjusted such that the onset masses $M_c$ for the deconfinement phase transition in a compact star lie at $2.0$, $1.39$ and $1.20$ M$_\odot$ for set 1, set 2 and set 3, respectively, while the maximum mass on the hybrid star branch exceeds the value of $2.01$ M$_\odot$ measured for PSR J0348+432 \cite{Antoniadis:2013pzd}. 
This is achieved by minimal variations in the low-density value $\eta_<$ of the vector coupling and the parameters of the switch functions $f_<(\mu)$ and $f_{\ll}(\mu)$ while keeping the bag constant $B=35$ MeV/fm$^3$ and $\eta_>=0.12$ constant.
{
See table \ref{tab:results} for the EoS parameters at the onset of the phase transition and the mass parameters characterizing the corresponding compact star sequences with a third family branch.
}

\begin{figure}[!ht]
	\includegraphics[width=0.6\textwidth]{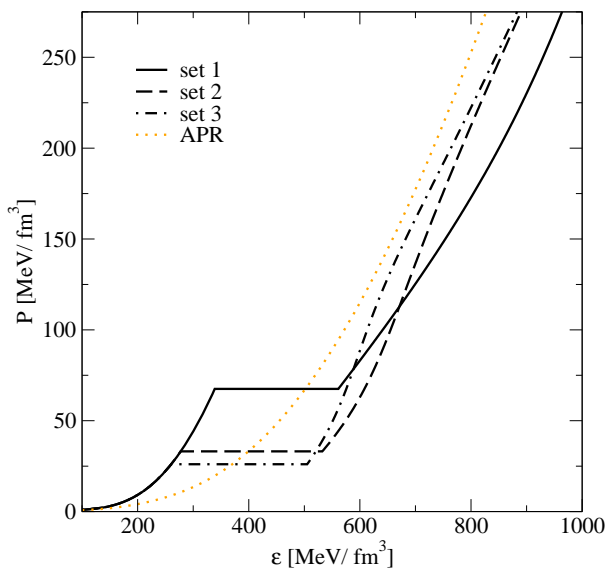}
	\caption{The equation of state for pressure vs. energy density resulting from the twofold interpolation method and Maxwell construction of the deconfinement phase transition illustrated in Fig.~\ref{fig:1a} for the parameters of set 2. Results for set 1, set 2 and set 3 are shown as solid, dashed and dash-dotted lines, respectively.
	For comparison, the APR EoS is shown (orange dotted line) which is a standard EoS for nuclear astrophysics applications}
	\label{fig:2}
\end{figure}

{
\subsection{Hybrid EoS and third family in  the M-R diagram}
}
In Fig.~\ref{fig:2} we show the hybrid EoS resulting from the Maxwell construction between the hadronic DD2\underline{ }p40 EoS and the quark matter EoS of sets 1, 2 and 3.

\begin{figure}[!ht]
	\includegraphics[width=0.6\textwidth]{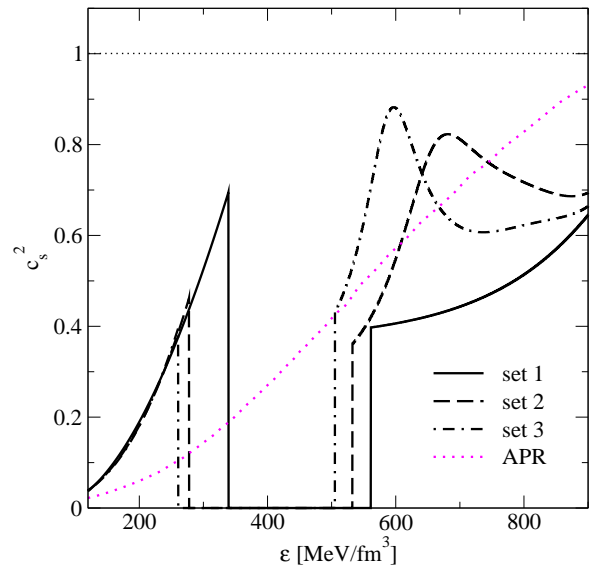}
	\vspace{-10mm}
	\caption{The squared speed of sound $c_s^2$ as a function of the energy density 
		for set 1, set 2 and set 3 exhibits the regions of the first order phase transition where $c_s^2=0$ and fulfills the condition of causality $c_s^2<1$ (in units of the speed of light squared). Line styles as in Fig.~\ref{fig:2}.
	For comparison, the APR EoS is shown (magenta dotted line) without deconfinement phase transition}
	\label{fig:3}
\end{figure}

Fig.~\ref{fig:3} shows the squared speed of sound $c_s^2=dP/d\varepsilon$ as a function of the energy density for sets 1, 2 and 3 which exhibit the regions of the first order phase transition where $c_s^2=0$ and fulfills the condition of causality $c_s^2<1$ (in units of the speed of light squared).

\begin{figure}[!th]
	\includegraphics[width=0.65\textwidth]{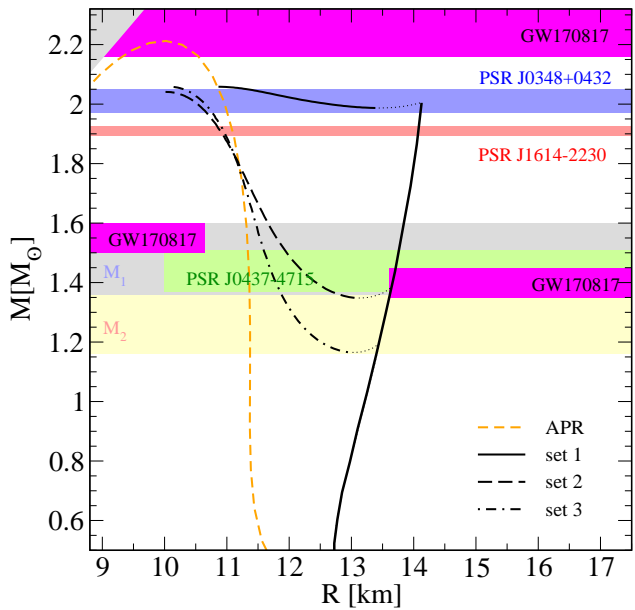}
	\vspace{-10mm}
	\caption{Mass vs. radius for sequences of compact stars with the hybrid EoS of this work parametrized by set 1, set 2 and set 3, corresponding to different onset masses for the deconfinement transition. The dotted lines denote the unstable configurations that should not be realized in nature but guide the eye to the corresponding stable hybrid star sequence (third family) disconnected from the neutron star one (second family). 
		For comparison, the mass-radius sequence for the APR EoS is shown (orange dashed line) which is a standard EoS for nuclear astrophysics applications, concerning only the second family, without a third family branch.
		The blue and red horizontal bands denote the mass measurement for PSR J0348+432 \cite{Antoniadis:2013pzd} and PSR J1614-2230 \cite{Arzoumanian:2017puf}, resp.
		The grey and orange bands labelled M1 and M2 are the mass ranges for the compact stars in the binary merger GW170817.
		From the observations of this event exclusion regions have been found (magenta hatching): Ref.~\cite{Bauswein:2017vtn} excludes radii of smaller than 10.68 km for stars of $1.6$ M$_\odot$ and Ref.~\cite{Annala:2017llu} excludes radii exceeding $13.6$ km for stars of 1.4 M$_\odot$. Ref.~\cite{Rezzolla:2017aly} derives an upper limit of 2.16 M$_\odot$ for the maximum mass of nonrotating neutron stars.
		The green band denotes the mass $1.44\pm 0.07$ M$_\odot$ of PSR J0437-4715, the primary target of the radius measurement by NICER \cite{Arzoumanian:2009qn}.}
	\label{fig:4}
\end{figure}

In Fig.~\ref{fig:4} we show a key result of this paper, the mass-radius relationships for the three hybrid EoS of Fig.~\ref{fig:2} as obtained from the solution of the TOV equations. 
The dotted lines denote the unstable configurations that should not be realized in nature but guide the eye to the corresponding stable hybrid star sequence (third family) disconnected from the neutron star one (second family). 
The blue and red horizontal bands denote the mass measurement for PSR J0348+432 \cite{Antoniadis:2013pzd} and PSR J1614-2230 \cite{Arzoumanian:2017puf}, resp.
The grey and orange bands labelled M1 and M2 are the mass ranges for the compact stars in the binary merger GW170817 for which Ref.~\cite{Bauswein:2017vtn} has excluded radii smaller than 10.68 km of $1.6$ M$_\odot$ stars and Ref.~\cite{Annala:2017llu} excludes radii exceeding $13.4$ km at 1.4 M$_\odot$. The green band denotes the mass range $1.44\pm 0.07$ M$_\odot$ of PSR J0437-4715, the primary target of the radius measurement by NICER \cite{Arzoumanian:2009qn}.
{
In the literature \cite{Rezzolla:2017aly,Shibata:2017xdx,Margalit:2017dij} a possible upper limit to the maximum mass of nonrotating compact stars has been deduced from the conjecture that GW170817 did not lead to a prompt black hole formation after the merger.
We indicate here the value $M_{\rm TOV}=2.16~$M$_\odot$ deduced in \cite{Rezzolla:2017aly}. 
}

\begin{figure*}[!htb]
\begin{center}$
\begin{array}{ccc}
\includegraphics[width=0.3\textwidth]{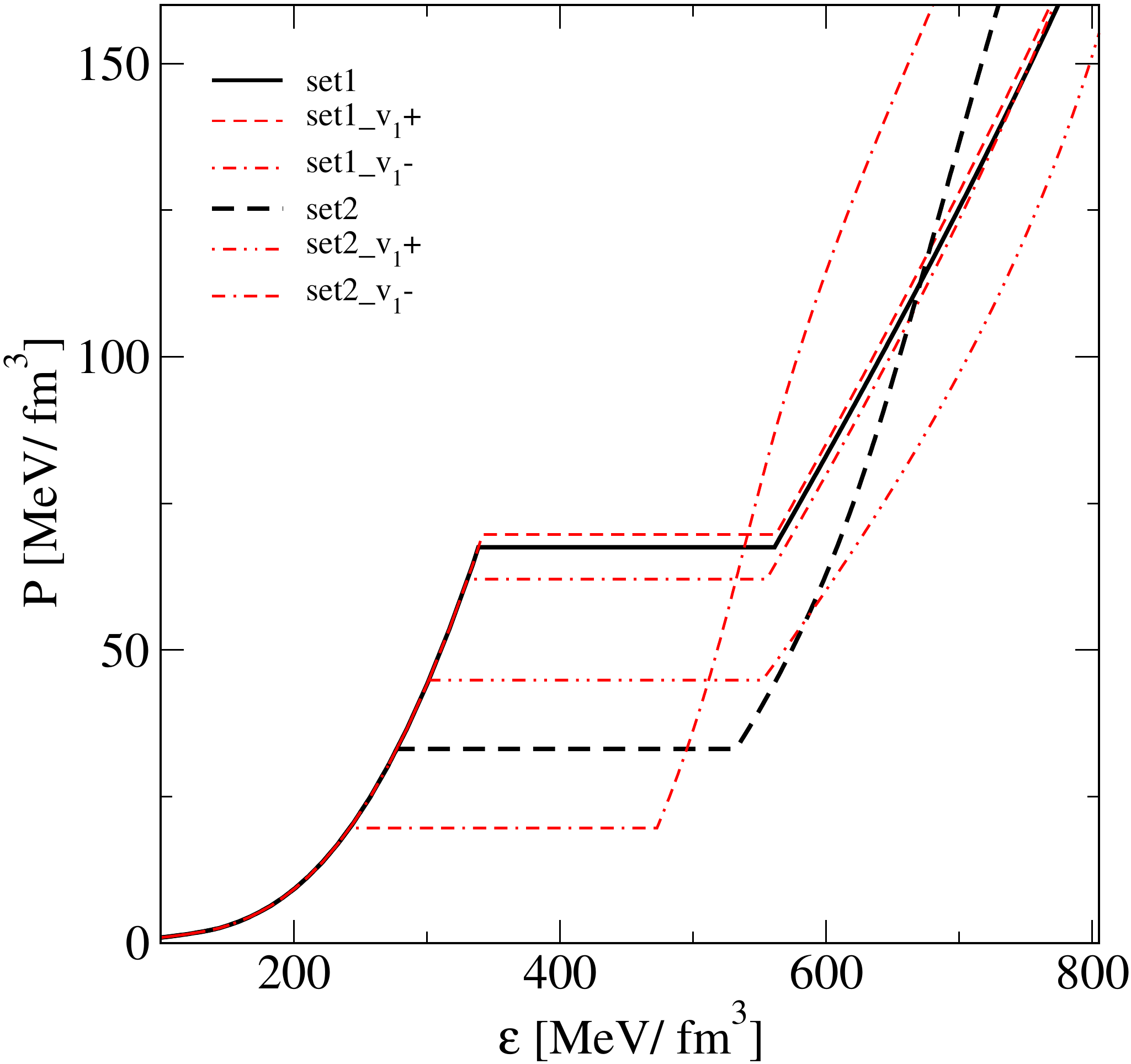} & 
\includegraphics[width=0.3\textwidth]{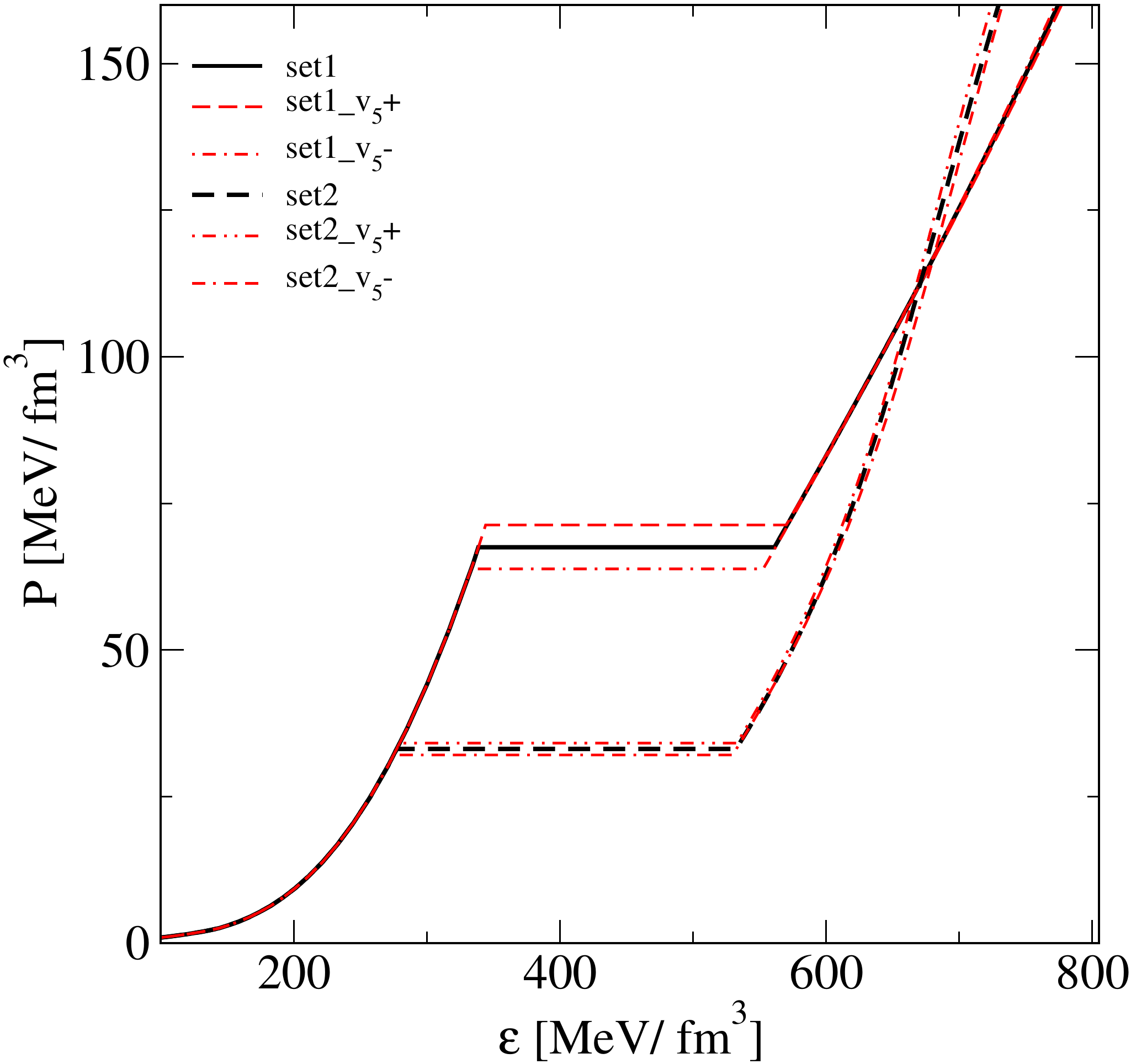} &
\includegraphics[width=0.3\textwidth]{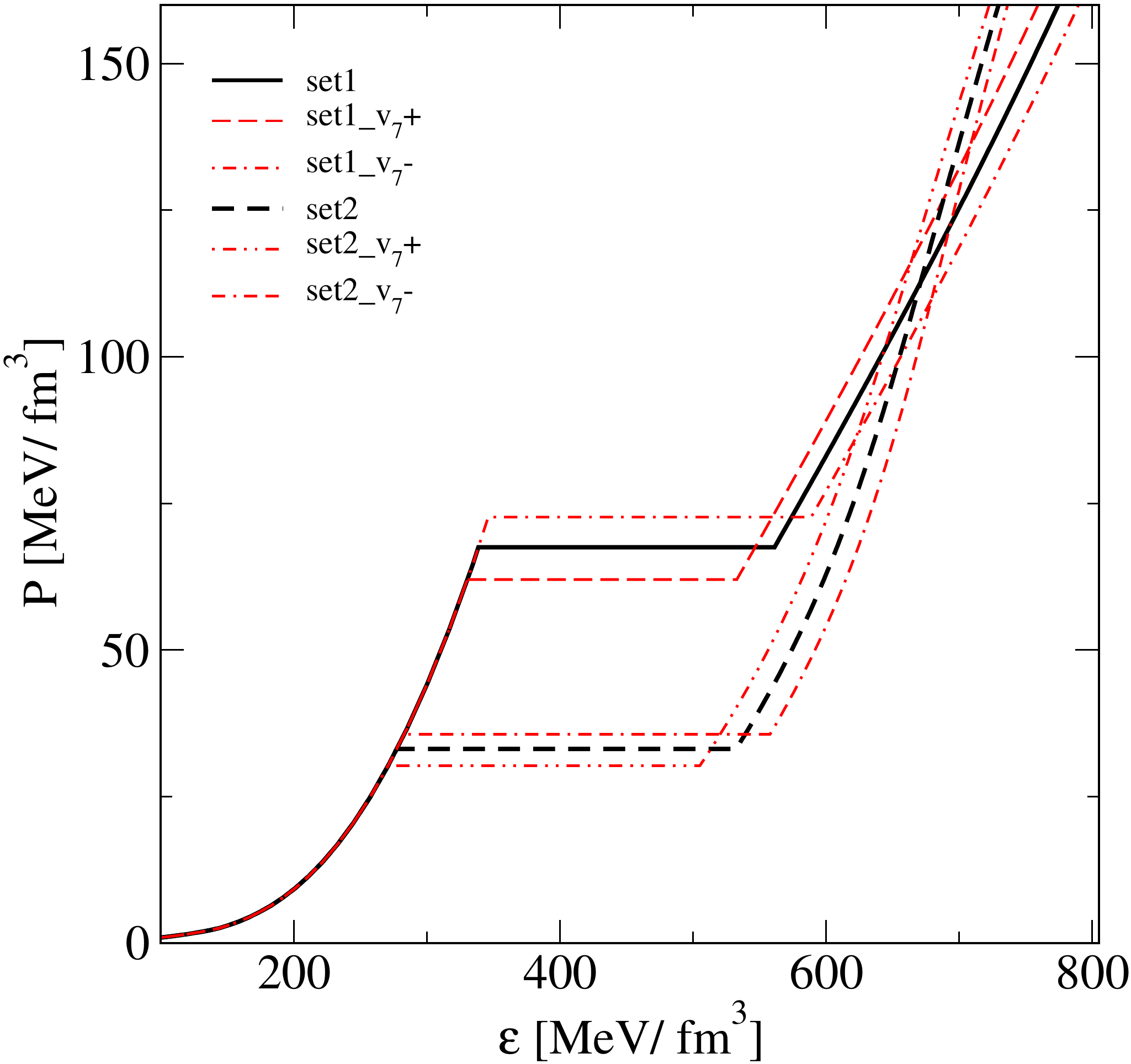} \\
\includegraphics[width=0.3\textwidth]{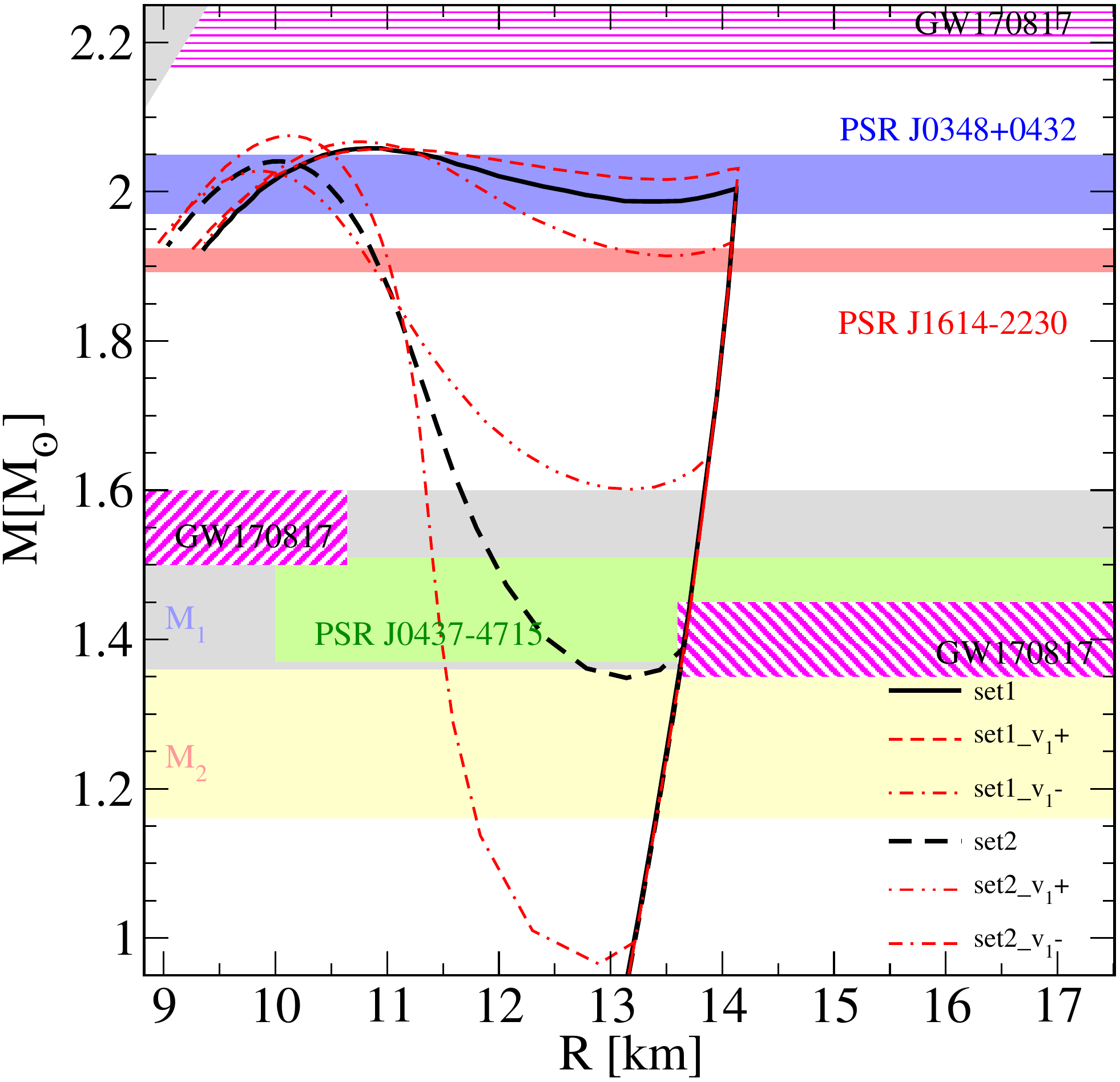}&
\includegraphics[width=0.3\textwidth]{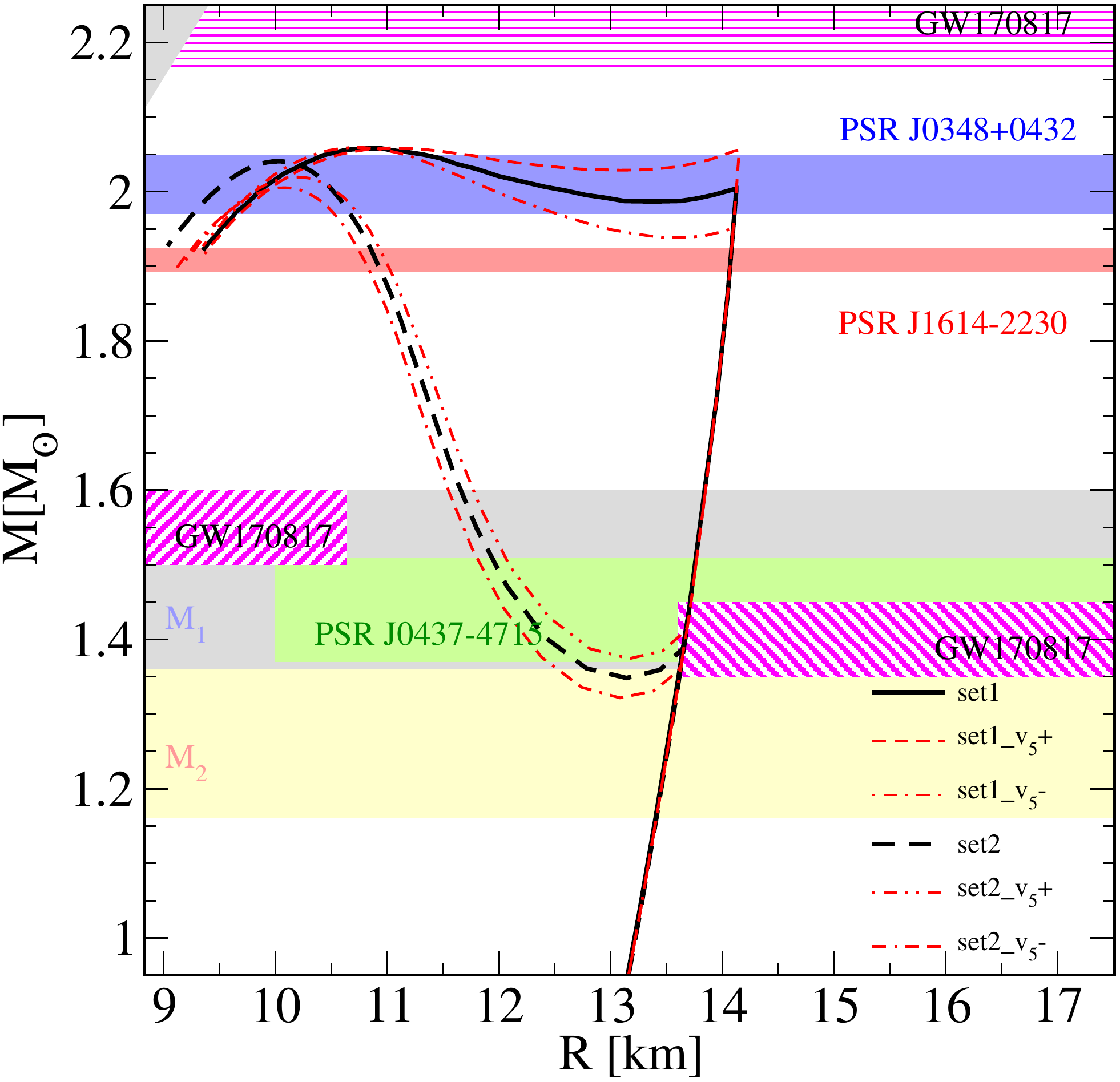}& 
\includegraphics[width=0.3\textwidth]{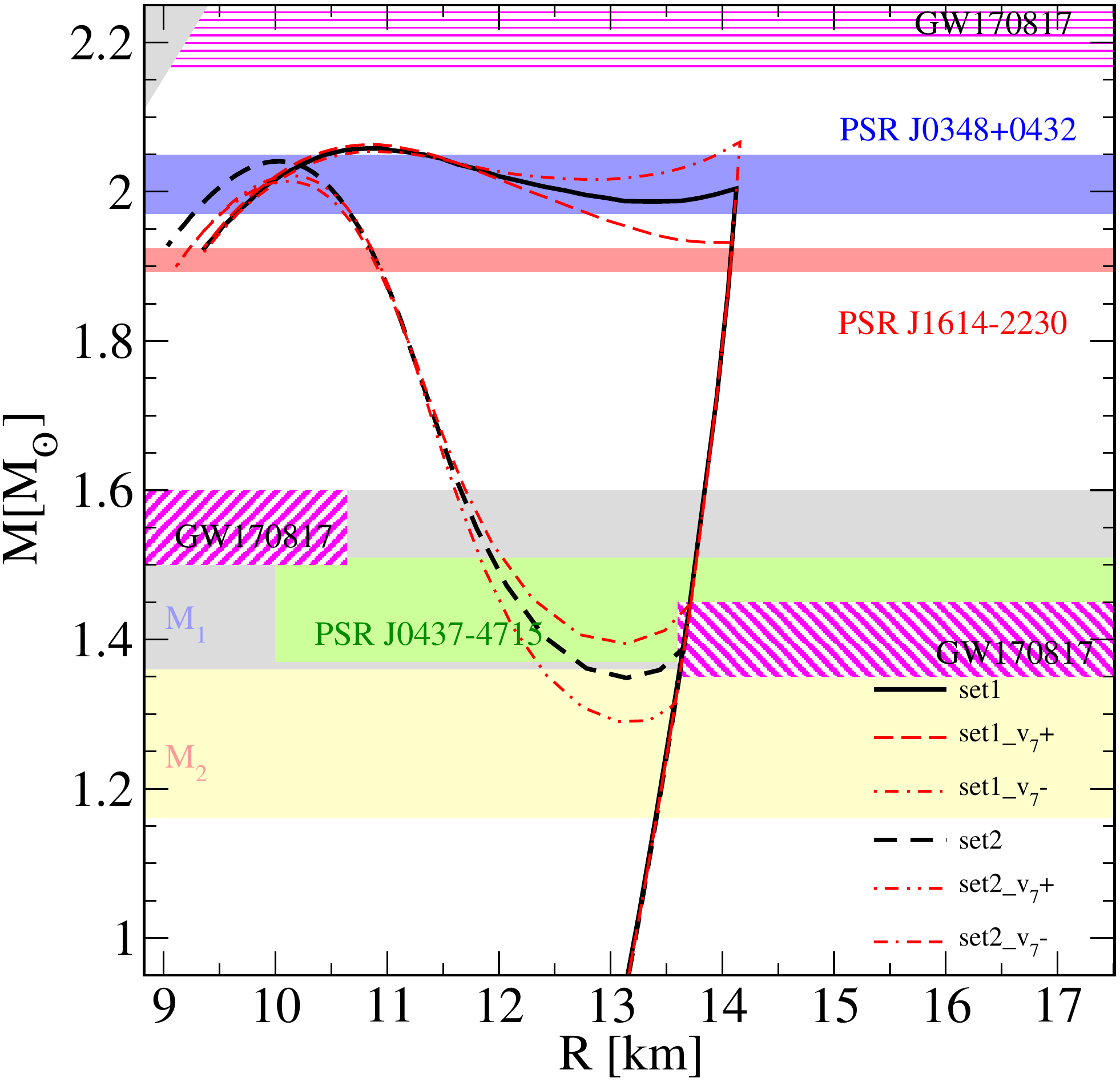}
\end{array}$
\end{center}
	\vspace{-5mm}
\caption{
\label{fig:varparlarge} Variation by $\pm 10\%$ of the interpolation parameters $v_1$ ($\mu_<$), $v_5$ ($\eta_<$) and $v_7$ ($B$), 
with a large effect on the EoS (upper row of panels)  and the $M-R$ sequences (lower row of panels) for set 1 (high onset mass) and set 2 (low onset mass). 
} 
\end{figure*}

\begin{figure*}[!htb]
\begin{center}$
\begin{array}{cc}
\includegraphics[width=0.35\textwidth]{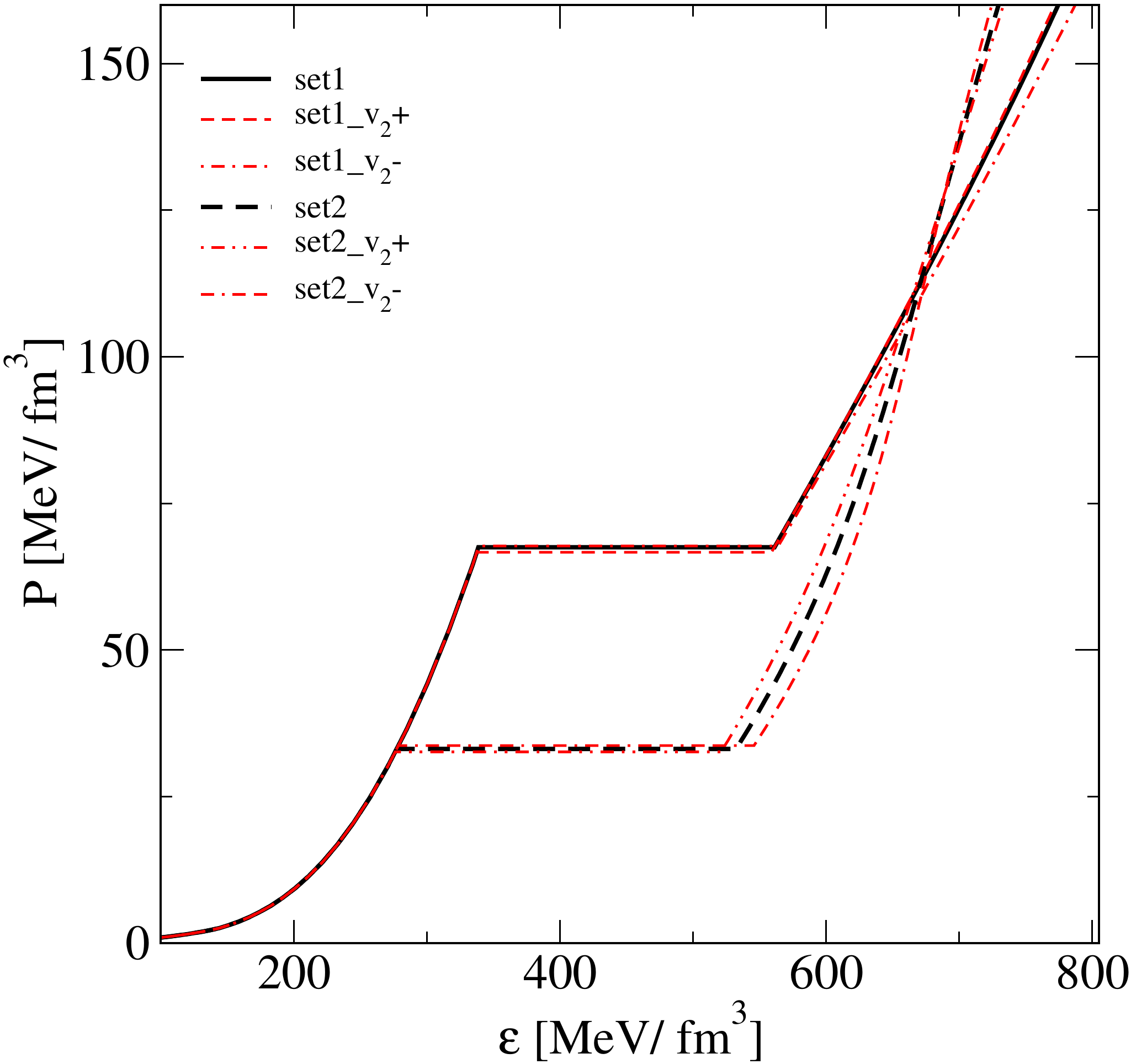} & \hspace{0cm}\includegraphics[width=0.35\textwidth]{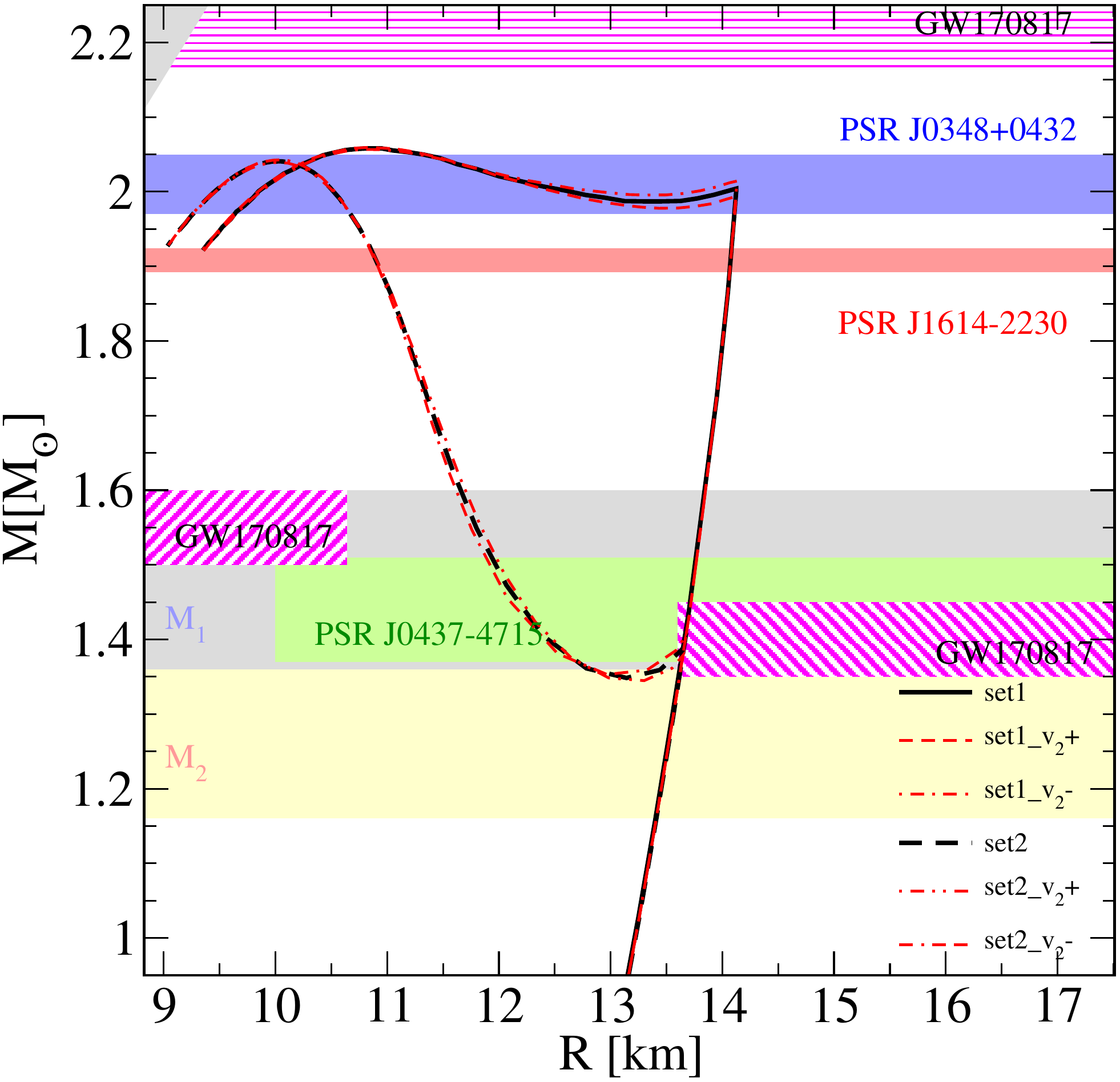}\\
\includegraphics[width=0.35\textwidth]{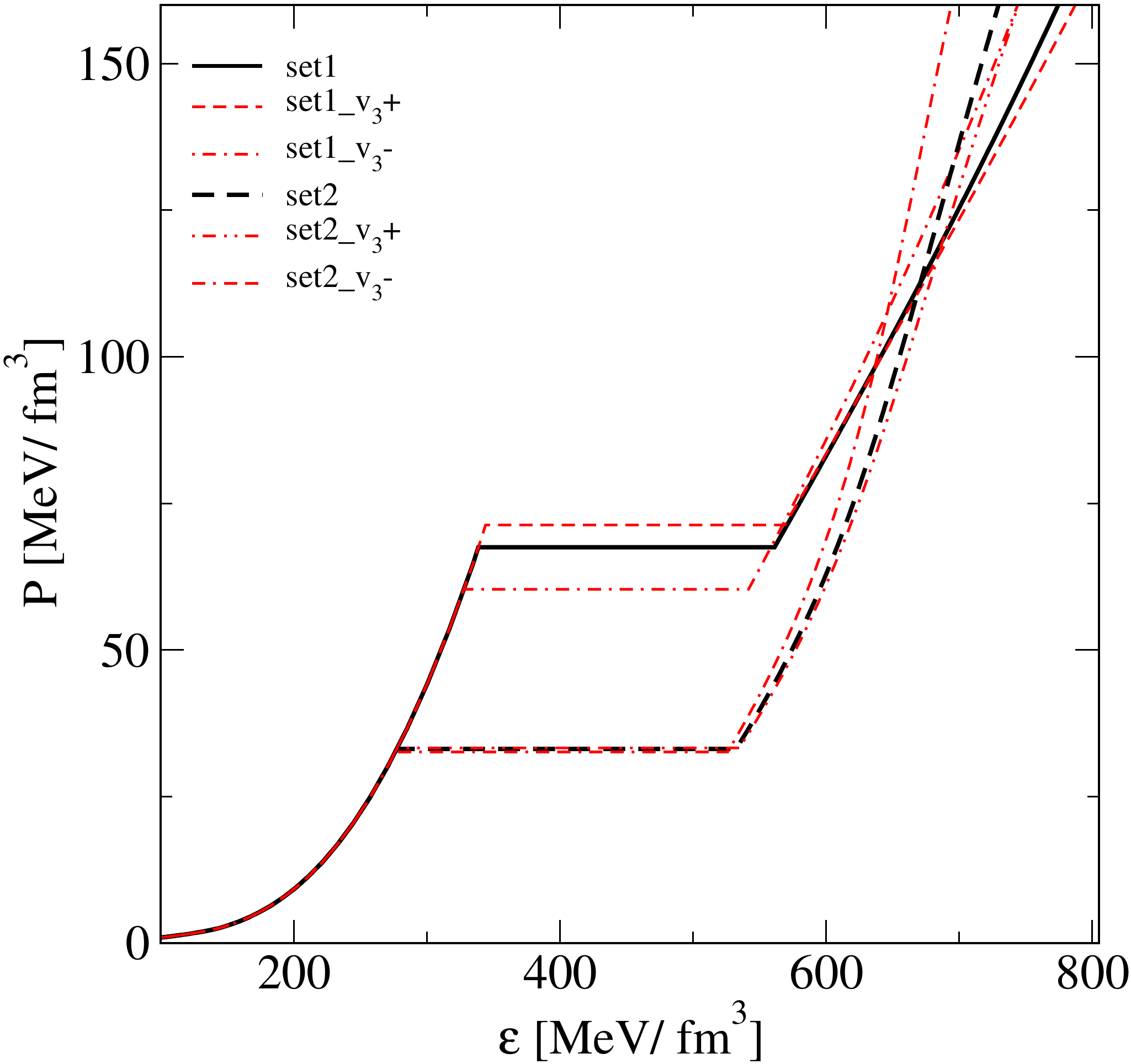} & \hspace{0cm}\includegraphics[width=0.35\textwidth]{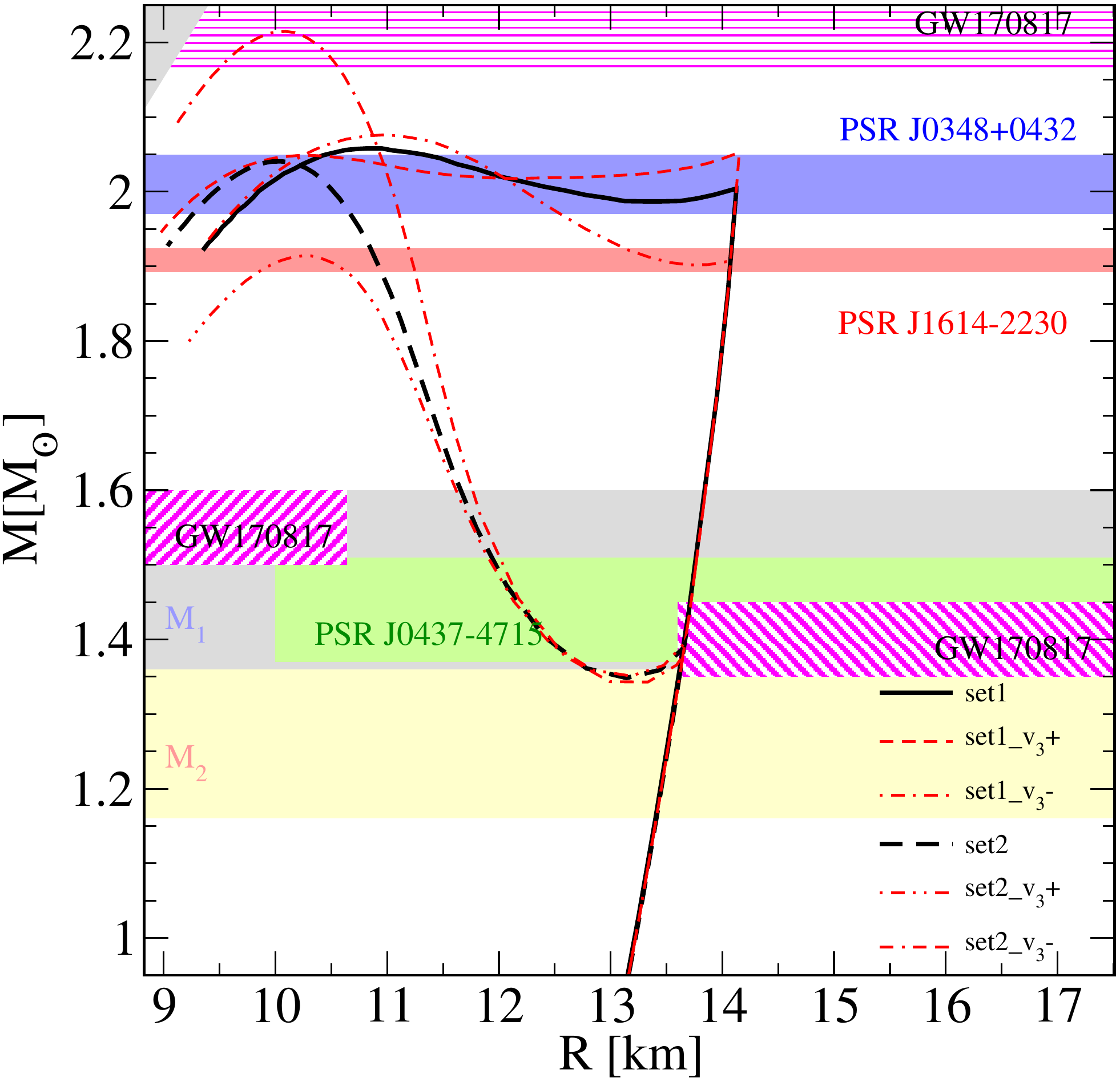}\\
\includegraphics[width=0.35\textwidth]{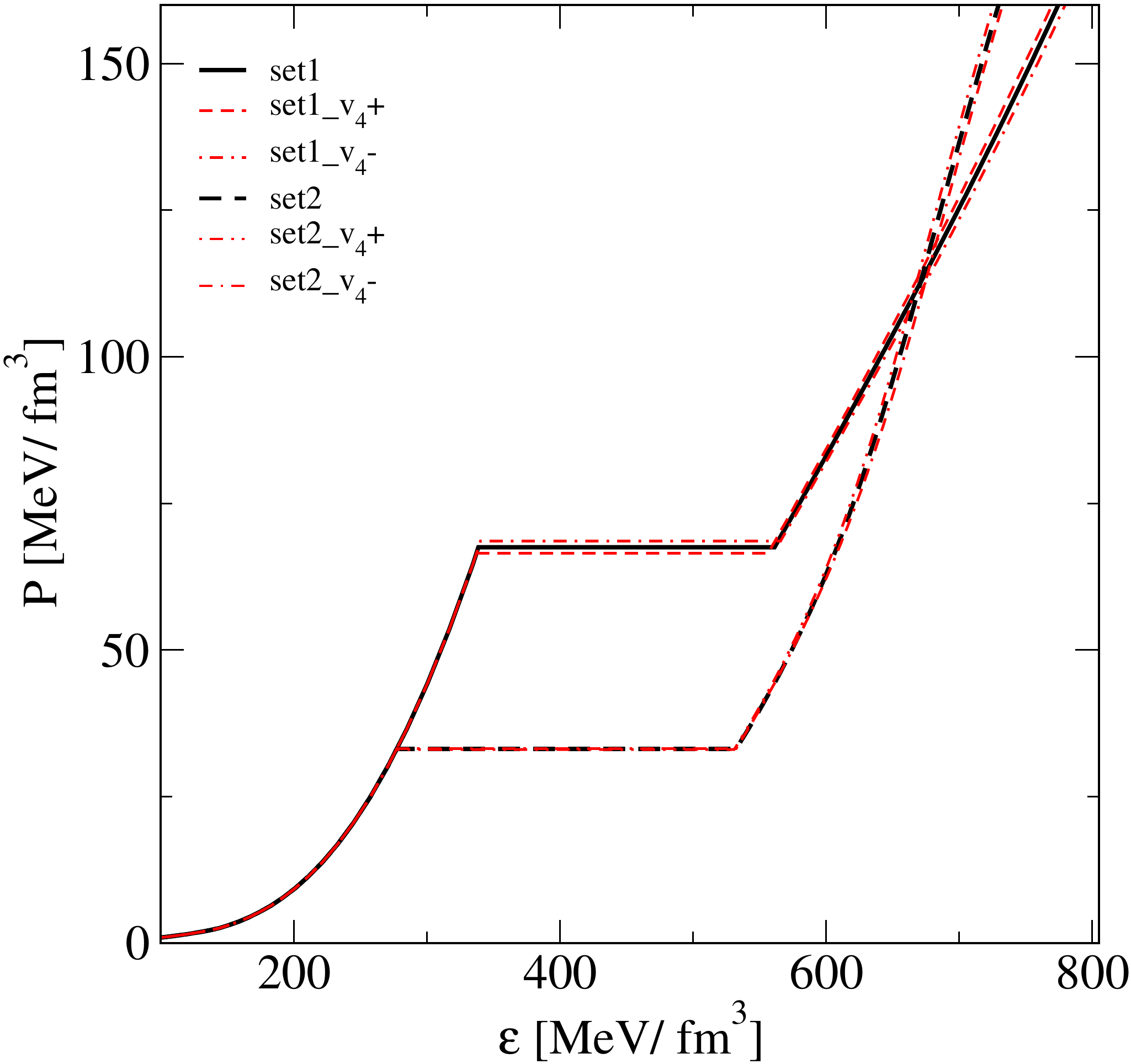} & \hspace{0cm}\includegraphics[width=0.35\textwidth]{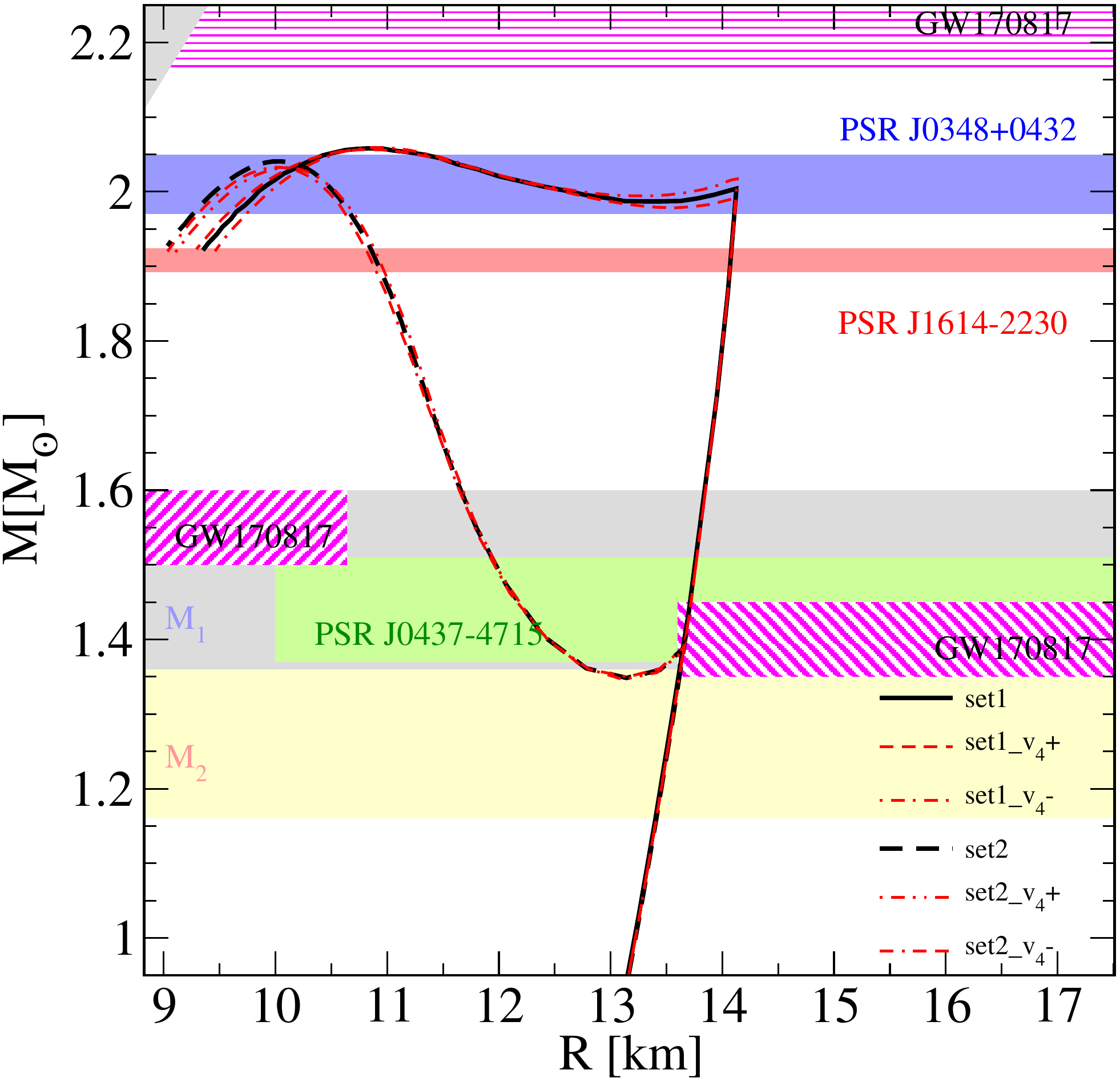}\\
\includegraphics[width=0.35\textwidth]{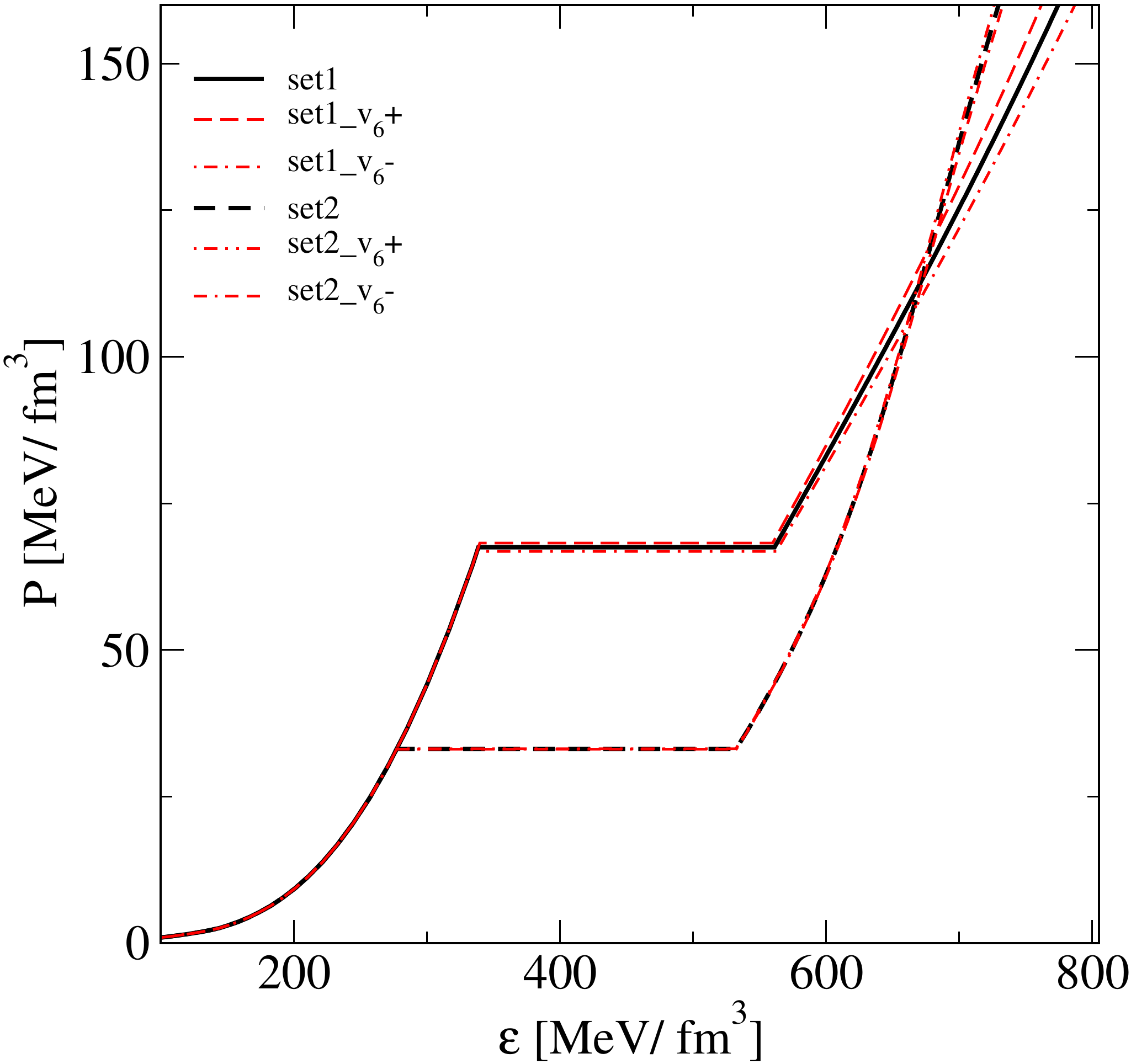} & \hspace{0cm}\includegraphics[width=0.35\textwidth]{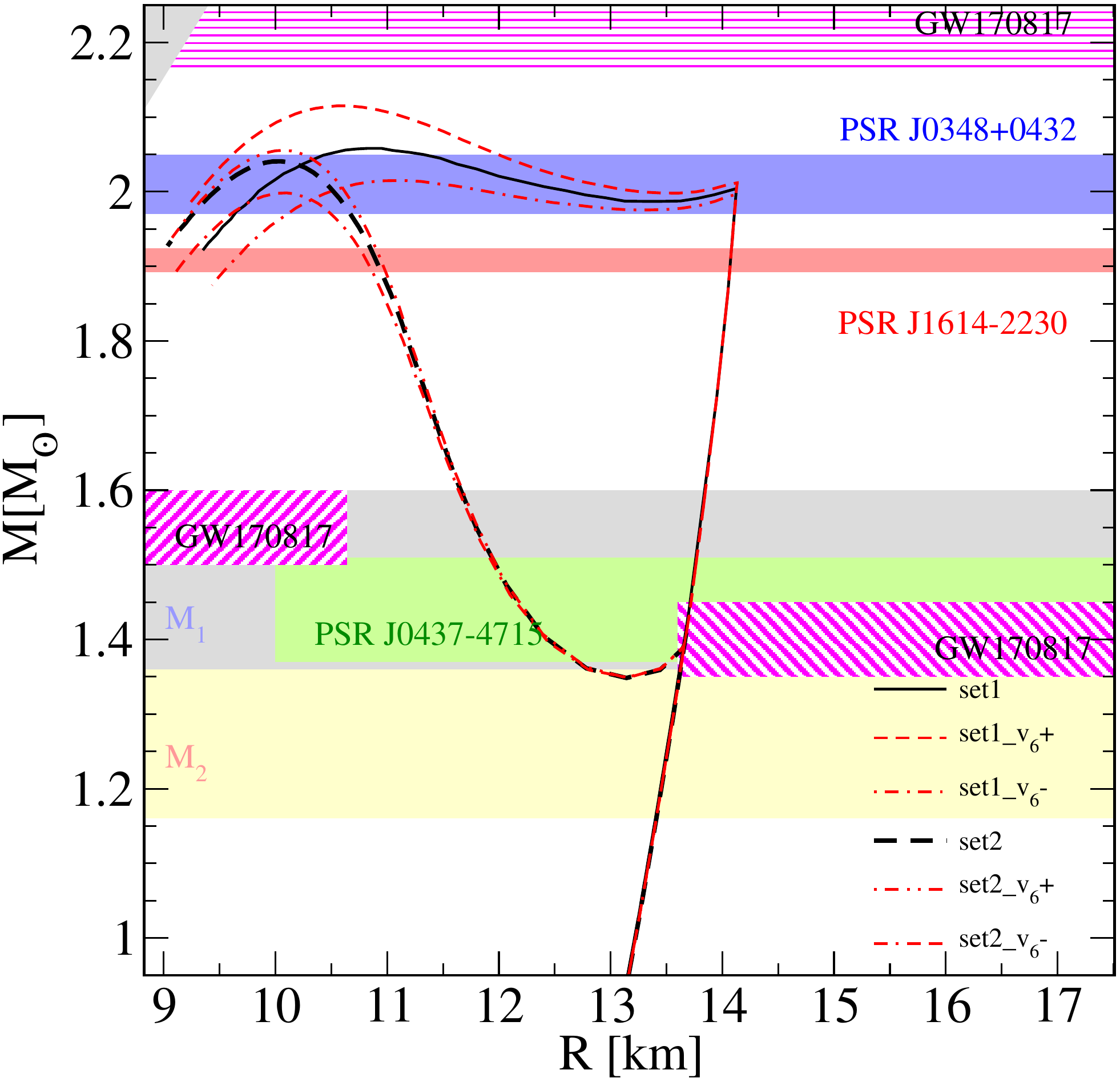}
\end{array}$
\end{center}
	\vspace{-5mm}
\caption{
\label{fig:varparsmall} Variation by $\pm 10\%$ of the interpolation parameters $v_2$ ($\Gamma_<$), $v_3$ ($\mu_\ll$), $v_4$ ($\Gamma_\ll$), and $v_6$ ($\eta_>$), 
with a smaller effect on the EoS (left column of panels)  and the corresponding $M-R$ sequences (right column of panels) for set 1 (high onset mass) and set 2 (low onset mass).
} 
\end{figure*}

{
\subsection{Sensitivity to interpolation parameters}

In this subsection we explore the sensitivity of the hybrid EoS for the generalized nonlocal NJL model against variations of the seven parameters of the twofold interpolation procedure given in table~\ref{tab:param}.
To this end we examine a $10\%$ increase and decrease of each of these parameters while keeping the others fixed.
The results for the EoS and mass-radius diagrams are shown in Figs.~\ref{fig:varparlarge} and \ref{fig:varparsmall}.
In this way we identify the onset of deconfinement $\mu_<$, the low-density vector coupling $\eta_<$ and the bag parameter $B$ as those parameters which have a sufficiently strong influence on the EoS to cause variations in the compact star sequences, see Fig.~\ref{fig:varparlarge}.
These parameters concern the low-density part of the quark matter EoS and thus the position of the deconfinement phase transition while the high-density part is less affected. 
Therefore the maximum mass of the hybrid star branch remains unaltered and fulfils the maximum mass constraint.

We note that this study allows to quantify the ambiguity in fixing the parameters of the interpolation method. 
For example, a $10\%$ increase in the bag parameter $B$ may be largely compensated by a $20\%$ decrease in the value of the low-density vector coupling $\eta_<$ or a $3\%$ reduction in the onset of deconfinement $\mu_<$, or a suitable combination of both.
Exploiting this knowledge we have generated the parameter set 3 with the aim to lower the onset mass for deconfinement to $1.2~M_\odot$ 
while keeping the third family branch and obeying the maximum mass constraint just by lowering the parameter $\mu_<$ for the onset of deconfinement by $5\%$ relative to set 2. 
}

{
We also identify those parameters to which the interpolation is less sensitive, see Fig.~\ref{fig:varparsmall}. 
The $\pm 10\%$ change of the width parameters for the interpolation of bag-melting ($\Gamma_<$) and vector coupling increase ($\Gamma_\ll$)
do not affect the results.

A change in the vector coupling constant at high densities $\eta_>$ affects only the high-density EoS and thus the maximum mass of the third family branch while leaving the onset mass of the deconfinement transition unchanged for both, set 1 and set 2.
Changing the location of the stiffening at high densities $\mu_\ll$ affects both, the maximum mass of the third family branch as well as the onset of deconfinement at high masses (set 2). 

It is worthwile to note that all those $\pm 10\%$ variations of the parameters in table~\ref{tab:param} do not spoil the existence of a third family and thus of twin star configurations.
}

{
\subsection{Further applications and their discussion}
}

\begin{figure}[!th]
	\includegraphics[width=0.6\textwidth]{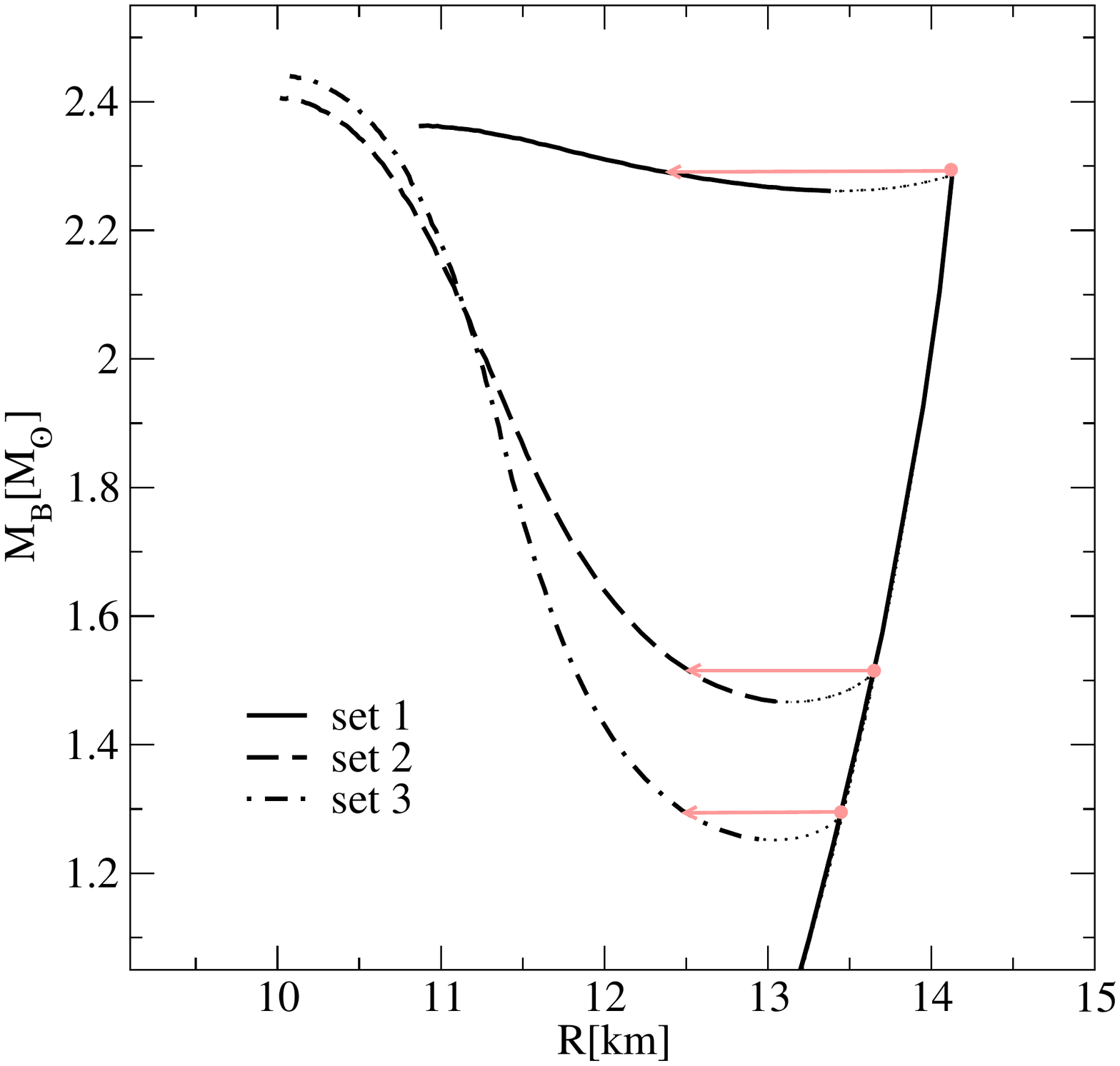}
	\vspace{-10mm}
	\caption{Baryon mass vs. radius for the compact star sequences corresponding to set 1, set 2 and set 3 parametrizations of the present EoS model. From this figure one can read off what drop in radius can be expected when a transition from the maximum mass of the second family branch to third family branch takes place under baryon number conservation, triggered for instance by mass accretion from a companion star (red arrows).}
	\label{fig:5}
\end{figure}

Fig.~\ref{fig:5} shows baryon mass vs. radius for the compact star sequences corresponding to set 1, set 2 and set 3 parametrizations of the present EoS model. 
From this figure one can read off what drop in radius can be expected when a transition from the maximum mass of the second family branch to third family branch takes place under baryon number conservation, triggered for instance by mass accretion from a companion star.

\begin{figure}[!th]
	\includegraphics[width=0.6\textwidth]{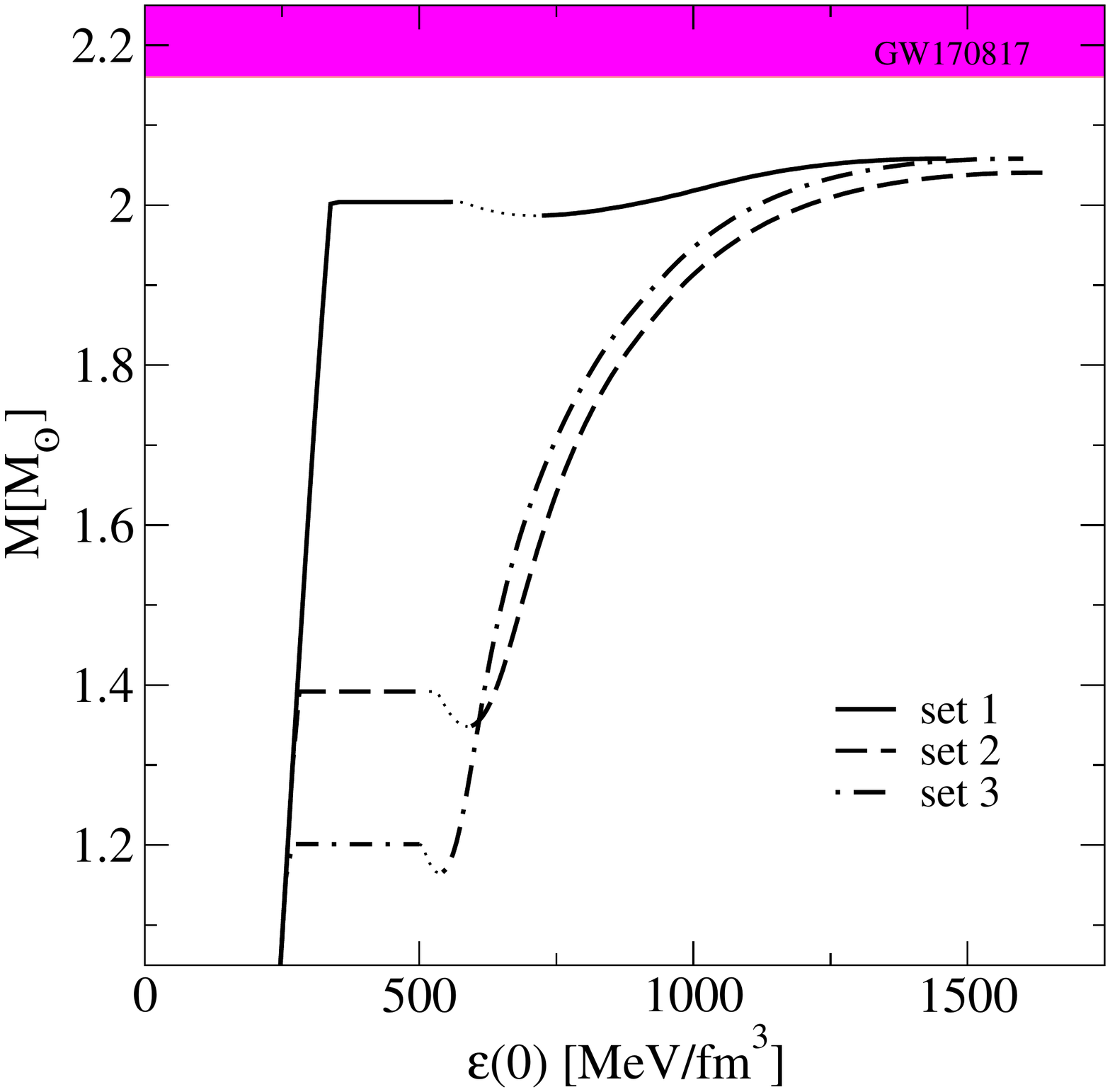}
	\vspace{-10mm}
	\caption{Mass vs. central energy density as solutions of the TOV equations for the three parameter sets given in Tab.~\ref{tab:param}, line styles as in the previous figures.
		The rising sections of the lines denote stable sequences.}
	\label{fig:6}
\end{figure}

In Fig.~\ref{fig:6} the mass vs. central energy density as solutions of the TOV equations for the three parameter sets given in Tab.~\ref{tab:param} is shown, line styles as in the previous figures. The rising sections of the lines denote stable sequences. 

\begin{figure}[!th]
	\includegraphics[width=0.6\textwidth]{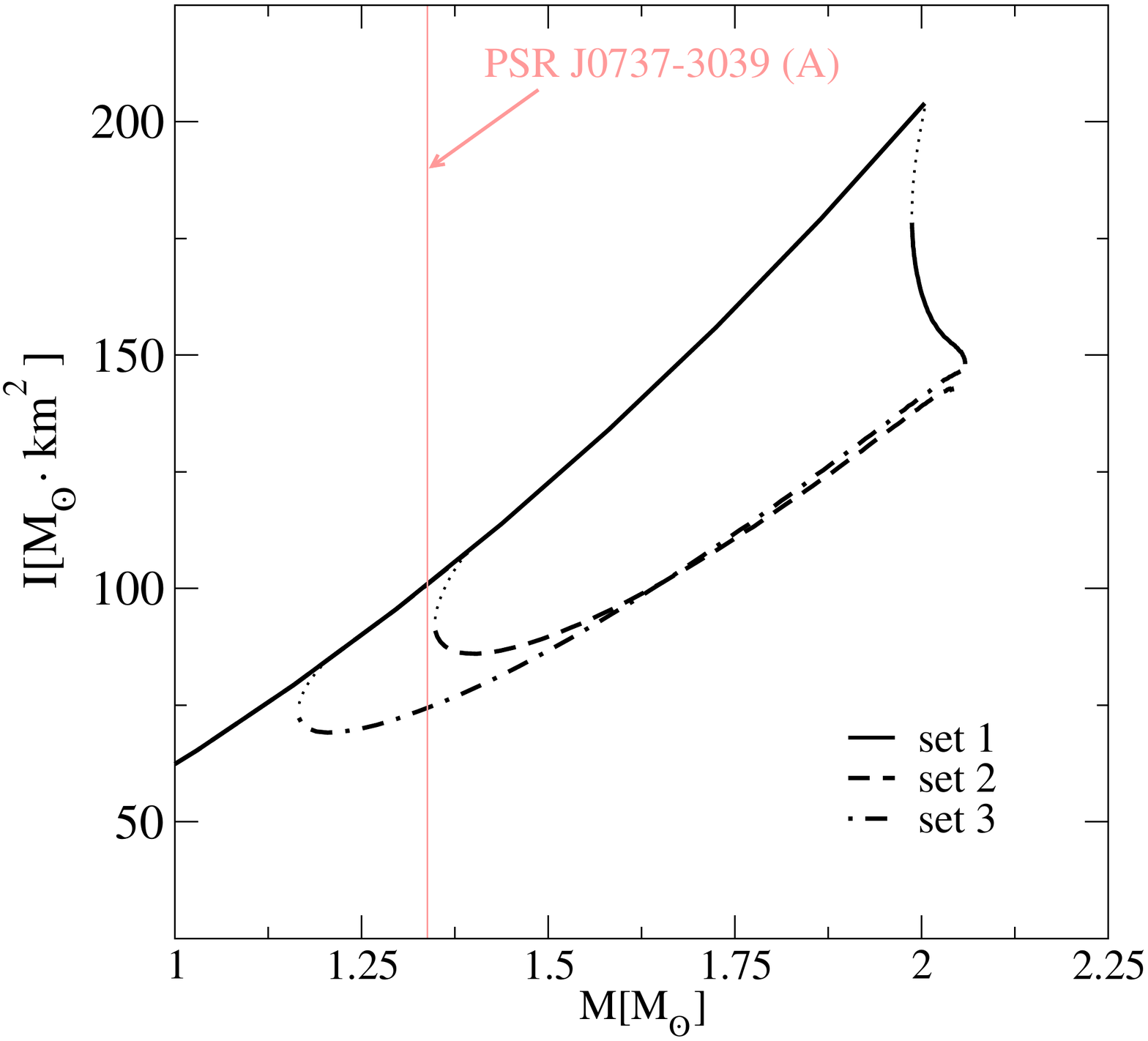}
	\vspace{-10mm}
	\caption{Moment of inertia vs. mass for the three hybrid EoS parametrizations introduced in this work. Unstable configurations are shown by dotted lines. Note that the relationship is multivalued in the mass twin ranges. Hybrid stars on the third family branch have generally a smaller moment of inertia. The mass of PSR J0737-3039 (A) is shown for which a measurement of I is expected at the 10\% level.}
	\label{fig:7}
\end{figure}

Fig.~\ref{fig:7} shows the moment of inertia vs. mass for the three hybrid EoS parametrizations introduced in this work. Unstable configurations are shown by dotted lines. Note that the relationship is multivalued in the mass twin ranges. Hybrid stars on the third family branch have generally a smaller moment of inertia.
Measurements of the moment of inertia of compact stars have the power to constrain the existing EoS models. Consider for instance the system PSR J0737-3039~\cite{Lyne:2004cj} where both compact stars A and B  have been observed as pulsars at the time of their discovery. Due to relativistic effects there exists a spin-orbit coupling that could eventually lead to the determination of $I_{A}$, the moment of inertia of star A in the binary. 
Since the masses of both A and B stars are already accurately determined to $M_A=1.3381(7)$ M$_\odot$ and  $M_B=1.2489(7)$ M$_\odot$ \cite{Kramer:2009zza}and the moment of inertia can be expressed in terms of only mass and radius and no other EoS parameters, the measurement of $I_{A}$ is of great importance as it allows for the simultaneous determination of mass and radius for the same object thus providing a datum of high discriminating power among all EoS models for compact stars~\cite{Lattimer:2004nj,Lattimer:2006xb}. 
Therefore, we show in Fig.~\ref{fig:7} the mass of PSR J0737-3039 (A) for which a measurement of $I_A$ at the prognosed 10\% level could well discriminate between a neutron star or hybrid star case in our example of the set 3 EoS parametrization. 

\begin{figure}[!th]
	\includegraphics[width=0.6\textwidth]{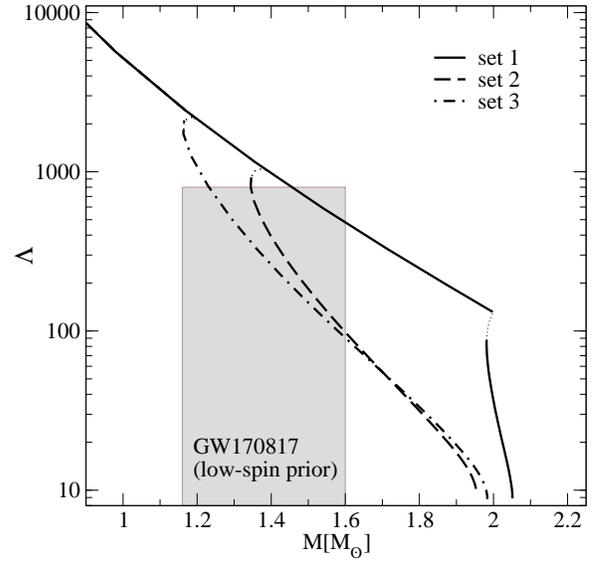}
	\vspace{-10mm}
	\caption{Dimensionless tidal deformability parameter $\Lambda$ vs. compact star mass for the three sets of hybrid EoS parametrizations introduced in this work.
		The constraint $\Lambda<800$ derived from GW170817 \cite{TheLIGOScientific:2017qsa}
		for the mass range 1.16 - 1.60 M$_\odot$ is fulfilled for the third family solutions 
		of set 2 and set 3 while the second family sequence of set 1 cannot fulfill this constraint within a binary neutron star scenario.}
	\label{fig:8}
\end{figure}

\begin{figure}[!th]
	\includegraphics[width=0.6\textwidth]{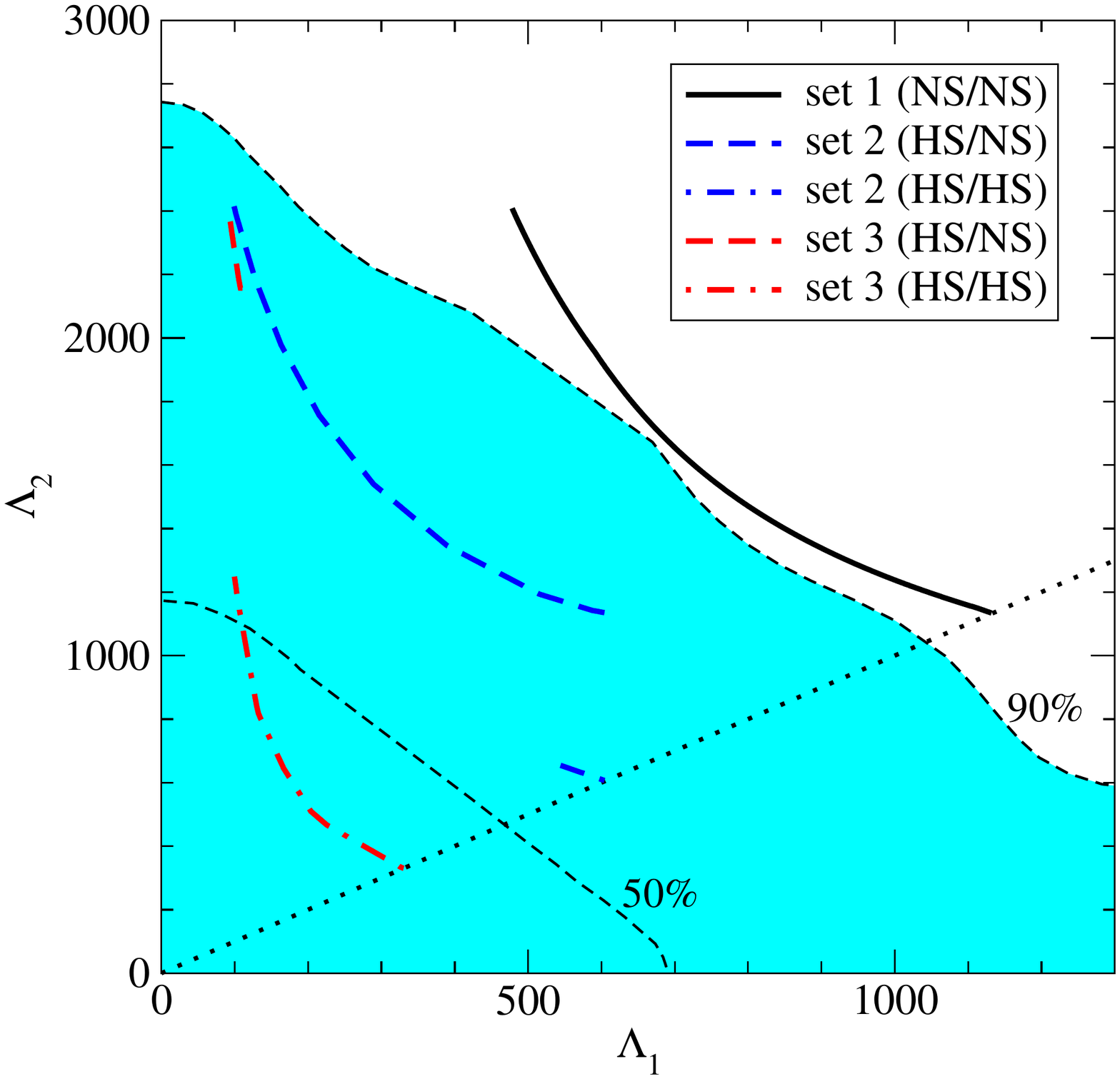}
	\vspace{-10mm}
	\caption{Tidal deformability parameters $\Lambda_1$ and $\Lambda_2$ of the high- and low-mass components of the binary merger, costructed from the $\Lambda(M)$ relation of Fig.~\ref{fig:8} for the EoS given by sets 1 - 3 compared to the probability contours for the low-spin prior of the LVC analysis of GW170817 \cite{TheLIGOScientific:2017qsa}.}
	\label{fig:9}
\end{figure}

In Fig.~\ref{fig:8} the dimensionless tidal deformability parameter $\Lambda$ vs. compact star mass for the three sets of hybrid EoS parametrizations introduced in this work.
The constraint $\Lambda<800$ derived from GW170817 \cite{TheLIGOScientific:2017qsa}
for the mass range 1.16 - 1.60 M$_\odot$ is fulfilled for the third family solutions 
of set 2 and set 3 while the second family sequence of set 1 cannot fulfill this constraint.

The second key result obtained for the new class of hybrid star EoS introduced in this work is shown in Fig.~\ref{fig:9}. It is the plot of the tidal deformability parameters $\Lambda_1$ and $\Lambda_2$ of the high- and low-mass components of the binary merger, constructed from the $\Lambda(M)$ relation of Fig.~\ref{fig:8} for the EoS given by sets 1 - 3 compared to the probability contours for the low-spin prior of the LVC analysis of GW170817 \cite{TheLIGOScientific:2017qsa}.
It shows that a stiff hadronic EoS like the one used in this work would be excluded by the tidal deformability constraint at the 90\% confidence level when it is to be used in a binary neutron star merger scenario. The same EoS being part of the three sets of hybrid EoS suggested in this work lead to acceptable scenarios when at least one of the stars in the coalescing binary belongs to the third family branch of hybrid stars. 
\\[5mm]

\section{Conclusions}

We have investigated the question whether it is possible to obtain a third family of compact stars for an equation of state resulting from the Maxwell construction of a first order phase transition between a relativistic meanfield EoS of nuclear matter (DD2F\underline{ }p40) and a nonlocal chiral quark model EoS of color superconducting two-flavor quark matter.
For the standard quark model with constant coupling strengths in the scalar, vector and diquark channels we find stable hybrid star branches that fulfill known constraints for masses and radii, however, being connected to the second family of purely nucleonic compact stars. 
{
No third family of compact stars is obtained in the standard nonlocal chiral quark model!
}

Therefore, we have applied in this work an interpolation procedure to { reconstruct} the thermodynamic behaviour of 
a class of hybrid compact star equations of state which supports the existence of a third family of compact stars \cite{Kaltenborn:2017hus}. 

To this end we had to generalize the nonlocal chiral quark model by addition of a confining, density-dependent bag pressure $B(\mu)$ 
{ 
that vanishes at high densities but at low densities leads to a sufficient softening of the EoS resulting in a large density jump at the 
deconfinement transition that triggers the occurrence of an unstable branch.
} 
In order to fulfill the constraint on the lower limit for the maximum mass of compact stars at $\sim 2$ M$_\odot$, the vector meson coupling 
$\eta(\mu)$ had to become a mildly increasing function of the chemical potential. 
For the chemical potential dependence of both, the vector coupling and the bag pressure, we have given analytic dependences on the smooth interpolation functions which themselves are well-constrained by causality and thermodynamic stability of the resulting quark matter EoS.

We have presented results for this class of hybrid equation of state and the corresponding properties of compact star sequences for three sets of parametrizations which result in a maximum mass of the second family (neutron star sequence) at 2.00, 1.39 and 1.20 $M_\odot$, respectively, and a third family of stable compact stars (hybrid star sequence) separated from the second one by a sequence of unstable configurations so that the phenomenon of mass twin stars is obtained. 
The hybrid star sequences include the observed maximum pulsar mass of $2.01$ M$_\odot$ as a necessary constraint for compact star EoS. 
{ 
While the sets 1 and 2 represent the reconstructed EoS of the string-flip model \cite{Kaltenborn:2017hus} the set 3 is an example of the class of EoS that can be generated within the generalized nonlocal NJL model and also corresponds to a third family branch of stable hybrid star configurations.

Despite the relatively large number of seven parameters in the twofold interpolation method we argue that it is worthwile to be defined in this way 
since these parameters have a physical meaning and thus allow an interpretation of the results of their fitting as well as their partial ambiguity.
We have investigated the sensitivity of the interpolation scheme against changes in the parameter values and demonstrated that only five parameters are relevant, but their relation to the compact star observables has a certain ambiguity that could be quantified.   
A parametrization like this shall become a powerful tool when applying it within a Bayesian analysis of observational constraints on masses and radii of compact stars \cite{Steiner:2010fz,Alvarez-Castillo:2016oln} as well as tidal deformabilities \cite{Ayriyan:2018blj} and further observables from compact star mergers \cite{Bauswein:2018bma}.
}

It is demonstrated that this advanced description of hybrid star matter allows to interpret GW170817 as a merger not only of two neutron stars but also of a neutron star with a hybrid star or of two hybrid stars. 
The latter two scenarios can be in accordance with the constraints on compactness from GW170817 when a binary neutron star merger scenario would be ruled out because of a too stiff hadronic equation of state. 
The NICER experiment on board of the International Space Station \cite{Arzoumanian:2009qn,Ray:2017tzb} has the potential to rule out too soft hadronic equations of state and thus support a merger scenario involving hybrid stars from a third family which can be described with the new class of EoS presented in this work.

\section*{Acknowledgements}

D.E.A-C. acknowledges support from JINR Dubna for participation in scientific events in Cuba, Mexico, South Africa and South Korea and for bilateral scientific collaboration with Institutes in Germany (Heisenberg--Landau programme) Poland (Bogoliubov--Infeld programme) and Armenia (Ter-Antonian--Smorodinsky programme).
This work has been supported in part by the Polish National Science Centre (NCN) under contract no. UMO-2014/13/B/ST9/02621 (D.E.A-C. and D.B.) and by the MEPhI Academic Excellence programme under contract no. 02.a03.21.0005 (D.B.).
We acknowledge the COST Actions CA15213 "THOR" and CA16214 "PHAROS" for supporting networking activities.

\end{document}